\documentclass[%
superscriptaddress,
 amsmath,amssymb,
]{revtex4-2}

\usepackage{graphicx}
\usepackage{bm}

\def\##1{\underline #1}
\def\=#1{\underline{\underline #1}}

\def\E{{\bf E}}
\def\e{{\bf e}}

\usepackage{hyperref}
\usepackage{cleveref}
\hypersetup{
    colorlinks=true,
    linkcolor=blue,
    citecolor=blue,
    urlcolor=blue,
    breaklinks=true,
}

\begin{document}
\preprint{APS/123-QED}

\title{Light propagation in time-periodic bi-isotropic media}

\author{Stefanos Fr.\ Koufidis}
\email{steven.koufidis20@imperial.ac.uk}
\affiliation{Blackett Laboratory, Department of Physics, Imperial College of Science, Technology and Medicine, Prince Consort Road, London SW7 2AZ, UK}
\author{Theodoros T.\ Koutserimpas}
\affiliation{School of Electrical and Computer Engineering, Cornell University, Ithaca, New York 14853, USA}
\author{Francesco Monticone}
\affiliation{School of Electrical and Computer Engineering, Cornell University, Ithaca, New York 14853, USA}
\author{Martin W.\ McCall}
\affiliation{Blackett Laboratory, Department of Physics, Imperial College of Science, Technology and Medicine, Prince Consort Road, London SW7 2AZ, UK}

\date{\today}

\begin{abstract}
Photonic structures and time-crystals, wherein time is incorporated as an additional degree of freedom for light manipulation, have necessitated the development of analytical and semi-analytical tools. However, such tools are currently limited to specific configurations, leaving several unexplored physical phenomena akin to photonic time-crystals elusive. In this communication, using a coupled-wave theory approach, we unveil the occurring light propagation phenomena in a time-periodic bi-isotropic medium whose permittivity, permeability, and chirality parameter are periodic functions of time. Contrary to their static counterparts, we demonstrate that the considered dynamic medium couples only co-handed counter-propagating waves. In cases of non-constant impedance, we prove that two first-order momentum gaps are formed in the Brillouin diagram, resulting in parametric amplification with different amplification factors and corresponding momenta for the right- and left-handed modes, respectively. The presence of chirality plays a major role in manipulating lightwave signals by controlling the center of resonance, the corresponding bandwidth, and the amplification factor in a distinct fashion for each mode. For a finite ``time-slab'' of the medium, we analytically derive the scattering coefficients as functions of time and momentum, discussing how extreme values of optical rotation grant access to the temporal analog of the chirality-induced negative refraction regime. Finally, we demonstrate the mechanism under which elliptical polarizations may change field orientation whilst the electric field propagates in a momentum gap, thus simultaneously showcasing parametric amplification.
\end{abstract}

\maketitle

\section{Introduction}
\label{Introduction}
\par When a lightwave encounters a spatial interface between two media with different electromagnetic properties, energy (i.e., frequency) is conserved, but momentum is not due to the broken translational invariance of space \cite{Joannopoulos2009}. On the contrary, as demonstrated in Morgenthaler's seminal work \cite{Morgenthaler1958}, when an electromagnetic wave encounters a temporal interface where the electromagnetic properties of the medium abruptly change at a certain moment, translational symmetry, and thus momentum, is conserved (see Noether’s theorem, e.g., in \cite{Zangwill2013}), while energy is not. One of the intriguing consequences of such frequency alteration is the occurrence of both forward and backward reflections \cite{Felsen1970, Fante1971, Fante1973}, which were recently experimentally observed in \cite{Moussa2023}.

\par While the notion of incorporating time as an additional degree of freedom has shown great promise in the realm of plasma physics, particularly in understanding how rapid ionization properties can affect the plasma permittivity \cite{Mendonça2000}, its direct application in light manipulation within optics and photonics have remained elusive for several decades. Nevertheless, the rapid expansion of photonic time-crystals \cite{Yao2017, Saha2023}, artificial meta-media \cite{Huidobro2021, Engheta2023}, as well as highly nonlinear materials \cite{Vezzoli2018, Bohn2020, Bruno2020, Lustig2023}, has provided the necessary framework for realizing exotic wave phenomena. These range from parametric amplification \cite{Koutserimpas2018}, traveling wave modulation \cite{Sotoodehfar2023}, dynamic spherical scattering \cite{Panagiotidis2023}, non-reciprocity \cite{Sounas2017}, and compensation of the lines of forces \cite{Pendry2021}, to broadband coherent wave control \cite{galiffi2023broadband}, unidirectional spectral effects \cite{Hayran2021}, enhanced absorption and impedance matching \cite{Li2021, Firestein2022, Hayran2023, Hayran2023b}Firestein2022, axion response in space-time crystals \cite{Prudencio2023}, and event cloaking \cite{McCall2010}, to name a few.

\par With maturing technology for time-modulation of a medium's parameters, a natural step forward is to explore the temporal modulation of electromagnetic chirality, as initially suggested in \cite{Mostafa2023}. In its well-studied spatial form, chirality results from two distinct mechanisms: magneto-electric coupling originating from molecular-scale mirror symmetry breaking and helical stacking of birefringent layers \cite{Arteaga2016, Wade2020}. Regarding the former, its hallmark effect is circular birefringence, whereby contra-handed circular polarizations propagate at different speeds \cite{Barron2009}. The temporal analog of such optical activity was first demonstrated in \cite{Yin2022}, and subsequently discussed in \cite{Mostafa2023, Mirmoosa2023arXiv}, wherein circular birefringence is manifested as two orthogonal circular polarizations with different frequencies for an abrupt temporal change of the medium's properties. On the other hand, the latter mechanism manifests as the circular Bragg phenomenon, whereby co-handed (with the medium's structural handedness) circularly polarized light is strongly backscattered whilst contra-handed light is transmitted \cite{Robbie1996, LakhtakiaBook2005, McCall2009}. The temporal analog of opalescence due to Bragg scattering in cholesteric liquid crystals was first investigated in \cite{Galiffi2022}, resulting in an ``Archimedes' screw'' capable of dragging and amplifying light. Within the same framework, \cite{Alavi2024} examined the case where the principal axes of an anisotropic permittivity tensor periodically wiggle.

\par By combining the coupled-wave theory description of \cite{Koutserimpas2022} with the Möbius transformation method of \cite{Koufidis2022b}, it was demonstrated in \cite{Koufidis2023b} that a periodic temporal modulation of a medium's permittivity results in a regime where parametric amplification occurs. For uniform modulation, analytic expressions were derived for the dispersion characteristics and the scattering coefficients, offering straightforward insights into the amplification mechanism. Inspired by the spatial version of periodic chiral structures, as seen in works such as \cite{Jaggard1989, Flood1994, Flood1995}, the present paper examines a medium in which the permittivity, the permeability, and the chirality are periodic functions of time. The focus is on how chirality influences the dispersion properties and optical response of the examined medium, demonstrating that chirality can indeed provide a unique opportunity to impart exotic properties to light propagation.

\par In particular, we demonstrate that for non-constant impedance, two first-order momentum gaps emerge: one corresponds to left-handed modes, whilst the other corresponds to the orthogonal mode. These two parametric amplification regimes are distinct, centered at different frequencies, with the dc-term of chirality controlling the locations of the central resonances. Additionally, both the dc- and ac-terms of chirality exert control over the resonances' bandwidths, as well as over the maximum achieved amplification. For a finite ``time-slab'' of the medium, if the values of the chirality are relatively small, the scattering coefficients reach levels similar to those of the achiral case. On the other hand, when the chirality is comparable to the medium's time-averaged refractive index, it becomes possible to access negative refracting states in the temporal domain. In these states, the direction of time-propagation and the temporal handedness of counter-propagating modes are interchanged. Lastly, we illustrate that the impact of chirality, even in cases where it assumes small values, is more pronounced on the rotation of the plane of polarization of a wave incident to a slab of the considered medium.

\par The manuscript is structured as follows: In Sec.\ \ref{Constitutive relations and the Beltrami fields}, we model the medium under consideration via the Condon constitutive relations. Due to homogeneity, by employing the Beltrami fields, we reduce the problem of electromagnetic wave propagation in such a bi-isotropic medium to that of wave propagation in a regular isotropic medium. In Sec.\ \ref{Coupled-wave theory description}, we introduce coupled-wave theory that aims to describe the wave propagation within the temporally-modulated complex medium via expressions that offer straightforward insights. In Sec.\ \ref{Dispersion characteristics}, we discuss the dispersion characteristics and the density of states, demonstrating the impact of chirality on the parametric amplification regime. In Sec.\ \ref{Propagation through a finite time-slab}, we examine the propagation of electromagnetic waves through a finite ``time-slab'' of the medium. We calculate the medium's response in terms of scattering coefficients and subsequently discuss the influence of giant chirality in the temporal analog of the negative refraction regime due to chirality. Moreover, we explore the dynamic evolution of the polarization state of an electromagnetic wave as it propagates through the slab. Finally, in Sec.\ \ref{Conclusions}, we summarize the impact of chirality on wave propagation in the examined medium.

\section{Constitutive relations and the Beltrami fields}
\label{Constitutive relations and the Beltrami fields}
\par When it comes to the simplest bi-isotropic reciprocal medium, characterized by a relative permittivity $\epsilon_r$, permeability $\mu_r$, and chirality parameter $g_r$, Maxwell's under-determined equations are typically supplemented by the so-called Tellegen constitutive relations \cite{BohrenBook2003}. In this study, we will be examining a medium whose relative parameters are periodic functions of time, sharing the same period $T>0$, i.e., $ f\left(t\right) = f\left(t+T\right)$, where $f=\{\epsilon_r, \mu_r, g_r\}\in\mathbb{R}$. The chosen modulation profiles are
\begin{equation}\label{modulation profile}
   f=\bar{f}+\delta f\frac{e^{i\Omega t}+e^{-i\Omega t}}{2}\,,
\end{equation}
where $\Omega={2\pi}/{T}$ is the temporal modulation frequency and $\delta f \ll 1$ represents the (weak) modulation strength of each parameter. The aforementioned connections are valid for \emph{instantaneous} responses; otherwise, they must be replaced by convolution integrals \cite{Mirmoosa2022}. Such instantaneous responses have been experimentally observed for all-optical modulation near the zero index of refraction regime \cite{Saha2023}. For example, in \cite{Tirole2023}, the response time of transparent conduction oxides (such as ITO \cite{Alam2016}) was found to be of the same order of magnitude as the cycle of radiation (a few femtoseconds). Furthermore, in \cite{Lustig2023}, aside from single-cycle response, high modulation depths were demonstrated ($\approx0.5$ times the real part of the refractive index).

\par While the Tellegen model has been successfully employed in \cite{Yin2022} to analyze the temporal analog of optical activity, at least as an approximation, it is essential to acknowledge that electromagnetic chirality
fundamentally arises from spatial dispersion \cite{Landau1984book}. Therefore, the Tellegen formalism  must be used with care. To address this concern, Mostafa {\it et al.}\ proposed in \cite{Mostafa2023} the utilization of the Condon model \cite{Condon1937} instead, which was demonstrated in \cite{Silverman1986} to be invariant under duality transformations, thus meeting the primary criterion for physically accepted constitutive relations. The primary rationale for employing the Condon model, in lieu of the Tellegen formalism, was that chirality could be introduced via the gyrotropic parameter, $g$, which is essentially non-dispersive \cite{Mostafa2023} when considered at frequencies substantially lower than the resonances of the chiral molecules (here, meta-atoms) \cite{SerdyukovBook2001}. 

\par In accordance with the notation of \cite{Koufidis2022a}, the symmetrized Condon model can be expressed as
\begin{equation}\label{Constitutive Relations}
    {\bf d}=\epsilon_r{\bf E}-gc_0\frac{\partial {\bf h}}{\partial t} \ {\rm and} \ {\bf b}=\mu_r{\bf h}+gc_0\frac{\partial {\bf E}}{\partial t}\,,
\end{equation}
where we have introduced the auxiliary fields ${\bf d}=\epsilon_0^{-1}{\bf D}$, ${\bf b}=(\eta_0/\mu_0){\bf B}$, and ${\bf h}=\eta_0{\bf H}$, which have the same dimensions as $\mathbf{E}$. Here, $\mathbf{E}$ and $\mathbf{B}$ are the complex-valued phasors of the primitive electromagnetic fields, while $\mathbf{D}$ and $\mathbf{H}$ are those of their associated stimulated excitation fields. Additionally, $\epsilon_0$, $\mu_0$, $c_0=1/(\epsilon_0\mu_0)^{1/2}$, and $\eta_0 =(\mu_0/\epsilon_0)^{1/2}$ denote the free-space permittivity, permeability, phase velocity of light, and impedance, respectively. The dimensionless chirality parameter is defined as $g_r = \omega c_0 g$, with $\omega$ being the angular frequency, ensuring clarity in distinguishing it from the conventional gyrotropy, such as the off-diagonal element of the permittivity tensor in the Faraday medium of \cite{Li2022}.

\par Due to the fact that circularly birefringent media possess an inherent circular basis, with circular polarizations being always eigenmodes of the isotropic case \cite{Gao2015}, and since all the constitutive parameters in Eq.\ \eqref{Constitutive Relations} are scalars \cite{Qiu2007}, we may utilize the so-called Beltrami fields. These are postulated as \cite{LakhtakiaBook1994}
\begin{equation}\label{Beltrami Fields}
{\bf E}_{\left(\pm\right)} = \frac{1}{2}\left({\bf E}\pm i\eta{\bf h}\right) \ {\rm and} \ {\bf h}_{\left(\pm\right)} = \frac{1}{2}\left[{\bf h}\mp \left({i}/{\eta}\right)\E\right]\,,
\end{equation}
where $\eta=\left({\mu_r}/{\epsilon_r}\right)^{{1}/{2}}$ and the subscript notation ``$+$'' (respectively, ``$-$'') denotes left- (respectively, right-) handed modes. Crucially, the fields of Eq.\ \eqref{Beltrami Fields} are self-dual to the actual electric and  magnetic excitation fields, respectively. This implies their invariance under duality transformations, as detailed in \cite{Lindell1992}. Consequently, these fields satisfy Maxwell's equations being self-dual to their corresponding actual fields, thus leaving Maxwell's equations invariant under duality transformations.

\par The ultimate purpose of this approach is to reduce the problem of wave propagation in a bi-isotropic medium to that of propagation in a regular \emph{isotropic} medium. This can indeed be achieved under two conditions, referred in Chap.\ 2 of \cite{Lindell1994book} as wavefield postulates (sic). The first requirement is that each of the wavefields, $\left({\bf E}_{\left(+\right)},{\bf h}_{\left(+\right)}\right)$ and $\left({\bf E}_{\left(-\right)},{\bf h}_{\left(-\right)}\right)$, will be ``experiencing'' an effective relative permittivity and permeability given by \cite{Yin2022, Mostafa2023}
\begin{equation}\label{Effective Permittivity and Permeability}
    \epsilon_{\left(\pm\right)}=\epsilon_r\left(1\pm \frac{g_r}{n_r}\right) \ {\rm and} \ \mu_{\left(\pm\right)}=\mu_r\left(1\pm \frac{g_r}{n_r}\right)\,,
\end{equation}
respectively, where $n_r=\left(\epsilon_r\mu_r\right)^{1/2}$. Consequently, the equivalent isotropic medium has constitutive relations given by \cite{Lindell1994book, Silverman1986}
\begin{equation}\label{Isotropic Constitutive Relations}
\left(\begin{matrix}{\bf d}_{\left(\pm\right)}\\{\bf b}_{\left(\pm\right)}\\\end{matrix}\right)=\left(\begin{matrix}\epsilon_{\left(\pm\right)}&0\\0&\mu_{\left(\pm\right)}\\\end{matrix}\right)\left(\begin{matrix}{\bf E}_{\left(\pm\right)}\\{\bf h}_{\left(\pm\right)}\\\end{matrix}\right)\,.
\end{equation}

\par The second requirement stipulates that if $\mathbf{E}$ and $\mathbf{h}$ satisfy Maxwell's equations, so are the individual wavefields of Eq.\ \eqref{Beltrami Fields}. Wherefore, the two wavefields are independent, i.e., decoupled. However, it is important to note that this condition can only be fulfilled in \emph{homogeneous} media, where the constitutive parameters remain constant throughout space. Given that we are dealing with a scenario involving global time modulation, which necessitates that at some moment, $t_0$, we start modulating the medium's parameters for \emph{all} points in space (at least for those traversed by the lightwave), the wavefields in Eq.\ \eqref{Beltrami Fields} will remain decoupled \cite{Yang2023}.

\par In the coupled-wave theory approach we take in this paper, we will be considering a sum of counter-propagating plane waves. This implies that the initial wave (before the start of the modulation) is already propagating and imparts a particular $k$-content, with $k$ being the wavenumber, that is conserved \cite{Galiffi2022photonics}. In our analysis, we will demonstrate how this wave evolves over time as the medium's parameters periodically change. It is worth noting that the initial wave may be anharmonic; hence, in reality, the medium may not need to be dynamically changing in infinite space but only in the space where the wave is propagating. 

\par The second postulate appears contradictory to the prevailing coupling mechanism in the spatial counterpart of the scenario under consideration, where strong coupling arises primarily between contra-handed counter-propagating modes \cite{Jaggard1989, Flood1994, Flood1995}. However, this apparent contradiction is expected due to the conservation of momentum, which is a characteristic feature of time-varying media \cite{Galiffi2022photonics, Ortega2023}, that forbids any chirality-reversing local reflections.

\section{Coupled-wave theory description}
\label{Coupled-wave theory description}
In a source-free medium, since the excitation ${\bf d}_{\left(\pm\right)}$-fields are solenoidal, combining Faraday's and Ampère-Maxwell's macroscopic curl relations, $\nabla\times\E_{\left(\pm\right)}=-c_0^{-1}{\partial {\bf b}_{\left(\pm\right)}}/{\partial t}$ and $\nabla\times{\bf h}_{\left(\pm\right)}=c_0^{-1}{\partial {\bf d}_{\left(\pm\right)}}/{\partial t}$, respectively, with the constitutive relations of Eq.\ \eqref{Isotropic Constitutive Relations}, we obtain
\begin{equation}\label{Wave Equation}
    c_0^2\nabla^2{\bf d}_{\left(\pm\right)}=\epsilon_{\left(\pm\right)}\frac{\partial\mu_{\left(\pm\right)}}{\partial t}\frac{\partial {\bf d}_{\left(\pm\right)}}{\partial t}+\epsilon_{\left(\pm\right)}\mu_{\left(\pm\right)}\frac{\partial^2 {\bf d}_{\left(\pm\right)}}{\partial t^2}\,.
\end{equation}
For a plane wave that is axially propagating along, say, the $z$-direction, since momentum is conserved, Eq.\ \eqref{Wave Equation} leads to two distinct equations via separation of the variables \cite{Koutserimpas2020}. Thus, we may decompose the excitation fields as ${\bf d}_{\left(\pm\right)}\left(z,t;k\right)={ \bf d}_{\left(\pm\right)}^{\rm s}\left(z;k\right){ d}_{\left(\pm\right)}^{\rm t}\left(t;k\right)$, where ${\bf d}_{\left(\pm\right)}^{\rm s}={d}_{\left(\pm\right)}^{\rm s}\hat{\bf e}_{\left(\pm\right)}$ and ${d}_{\left(\pm\right)}^{\rm t}$ denote the (vectorial) spatial and the (scalar) temporal parts of the field, respectively. Since the wavefields decomposition splits the power carried by the wave into two independent orthogonal circular polarizations \cite{Lindell1992}, we have
\begin{equation}\label{Polarization Vector}
    \hat{\bf e}_{\left(\pm\right)}=\frac{1}{\sqrt{2}}\left(\begin{matrix}1\\\pm i\\\end{matrix}\right)\,,
\end{equation}
where ``$+$'' (respectively, ``$-$'') indicates left- (respectively, right-) handed circular polarization \cite{Hecht2017}. 

\par Whence, upon substituting the decomposed field into Eq.\ \eqref{Wave Equation} and separating the variables, with $-k^2$ being the separation constant, we arrive at two equations. Specifically, for the spatial part of the field, Eq.\ \eqref{Wave Equation} leads to the usual Helmholtz wave equation
    \begin{equation}\label{Wave Equation Spatial Part}
     \frac{{\rm d^2}{\bf d}_{\left(\pm\right)}^{\rm s}}{{\rm d} z^2}+k^2{\bf d}_{\left(\pm\right)}^{\rm s}={\bf 0}\,,
\end{equation}
whereby assuming that both wavefields propagate in the positive (spatial) direction, the solutions to Eq.\ \eqref{Wave Equation Spatial Part} are
\begin{equation}\label{Solution for the Spatial Part}
    {\bf d}_{\left(\pm\right)}^{\rm s}=d_{\left(\pm\right)}^{\rm s}e^{ikz}\hat{\bf e}_{\left(\pm\right)}\,.
\end{equation}
Accordingly, for the temporal part, Eq.\ \eqref{Wave Equation} leads to 
\begin{equation}\label{Wave Equation Temporal Part}
   \frac{{\rm d}^2 d_{\left(\pm\right)}^{\rm t}}{{\rm d} t^2}+\left[\theta_1\right]_{\left(\pm\right)}  \frac{{\rm d} d_{\left(\pm\right)}^{\rm t}}{{\rm d} t}+\left[\theta_2\right]_{\left(\pm\right)}   d_{\left(\pm\right)}^{\rm t}=0\,,
\end{equation}
where we have compactly written
\begin{equation}\label{Theta 1 and 2}
    \left[\theta_1\right]_{\left(\pm\right)}  = \frac{1}{\mu_{\left(\pm\right)}}\frac{{\rm d} \mu_{\left(\pm\right)}}{{\rm d} t} \ {\rm and} \ \left[\theta_2\right]_{\left(\pm\right)}  =\frac{k^2c_0^2}{\epsilon_{\left(\pm\right)}\mu_{\left(\pm\right)}}\,.
\end{equation}

\par As reviewed in Appx.\ \ref{Fourier expansions of the characteristic coefficients}, taking the Fourier expansions of the functions in Eq.\ \eqref{Theta 1 and 2}, 
\begin{equation*}
     \left[\theta_{1,2}\right]_{\left(\pm\right)}  =\sum_{n=0}^{+\infty}{ \left[\tilde{\theta}_{1,2}\right]_{\left(\pm\right)}^n\cos{\left(n\Omega t\right)}}\,,
\end{equation*}
the retained terms for weak modulation amplitudes are 
\begin{equation}\label{Fourier Expansion}
     \left[\theta_{1,2}\right]_{\left(\pm\right)}\approx \left[\bar{\theta}_{1,2}\right]_{\left(\pm\right)}+\left[\delta\theta_{1,2}\right]_{\left(\pm\right)}\frac{e^{i\Omega t}\mp e^{-i\Omega t}}{2}\,.
\end{equation}
For the $\left[\theta_1\right]_{\left(\pm\right)}$ coefficients, the dc-terms and the amplitudes of the modulation depths are found to be
\begin{subequations}
  \begin{equation}\label{dc and amplitude theta 1}
   \left[\bar{\theta}_{1}\right]_{\left(\pm\right)}=0 \ {\rm and} \ \left[\delta\theta_{1}\right]_{\left(\pm\right)}=i\Omega\left(\frac{\delta \mu_r}{\bar{\mu}_r}+\frac{\delta n_{r,\left(\pm\right)}}{\bar{n}_{r,\left(\pm\right)}}-\frac{\delta n_r}{\bar{n}_r}\right)\,,
\end{equation}
respectively, whereas for the $\left[\theta_2\right]_{\left(\pm\right)}$ coefficients the equivalent terms are
\begin{equation}\label{dc and amplitude theta 2}
   \left[\bar{\theta}_{2}\right]_{\left(\pm\right)}=\frac{k^2c_0^2}{\bar{n}_{r,\left(\pm\right)}^2} \ {\rm and} \ \left[\delta\theta_{2}\right]_{\left(\pm\right)}= -2k^2c_0^2\frac{\delta n_{r,\left(\pm\right)}}{\bar{n}_{r,\left(\pm\right)}^3}\,,
\end{equation}  
\end{subequations}
respectively; for brevity, the meaning of each symbol is explicitly defined in Appx.\ \ref{Fourier expansions of the characteristic coefficients}, but the notation is clear: barred values correspond to dc-terms, while the $\delta$-symbols correspond to first-order modulation strengths.

\par It is easily understood from Eqs.\ \eqref{Fourier Expansion} and \eqref{dc and amplitude theta 1} that under the assumption of weak modulation, if we set $t_0=0$, then $\left[\theta_{1}\right]_{\left(\pm\right)}\left(t_0\right)=0$. These observations naturally guide us to the field transformation \cite{Hartman1982}
\begin{equation}\label{Field Transformation}
  {y}^{\rm t}_{\left(\pm\right)} = {\rm exp}\left({\frac{1}{2}\int_{t_{0}}^t {{{\left[ {{\theta _1}} \right]}_{\left(\pm\right)}}{\rm d}t'} }\right)d_{\left(\pm \right)}^{\rm t}\,,
\end{equation}
that renders Eqs.\ \eqref{Wave Equation Temporal Part} as two Hill equations, namely
\begin{equation}\label{Hill Equation}
  \frac{{{{\rm{d}}^2}{{y}_{\left(\pm\right)}}}}{{{\rm{d}}{t^2}}} + \left[\theta\right]_{\left(\pm\right)}{{y}_{\left(\pm\right)}} = 0\,.
\end{equation}
The identified coefficients in Eq.\ \eqref{Hill Equation} are
\begin{equation}\label{Theta Coefficient}
    \left[\theta\right]_{\left(\pm\right)}={{{\left[ {{\theta _2}} \right]}_{\left(\pm\right)}} - \frac{1}{2}\frac{{{\rm{d}}{{\left[ {{\theta _1}} \right]}_{\left(\pm\right)}}}}{{{\rm{d}}t}} - \frac{1}{4}\left[ {{\theta _1}} \right]_{\left(\pm\right)}^2}\,,
\end{equation}
with Fourier components, as per the $\left[\theta_2\right]_{\left(\pm\right)}$ expansions in Eq.\ \eqref{Fourier Expansion} (n.b.\ the minus sign between the exponentials),
\begin{align}\label{dc and amplitude theta}
     \left[\bar{\theta}\right]_{\left(\pm\right)}=&   \left[\bar{\theta}_{2}\right]_{\left(\pm\right)}+\frac{1}{8}\left[\delta\theta_{1}\right]_{\left(\pm\right)}^2\,, \nonumber \\
     \left[\delta\theta\right]_{\left(\pm\right)}=& \left[\delta\theta_2\right]_{\left(\pm\right)}-i\left({\Omega}/{2}\right)\left[\delta\theta_1\right]_{\left(\pm\right)}\,.
\end{align}

\par Analogously to the application of coupled-wave theory in spatially modulated dielectric media, such as structurally chiral media \cite{Koufidis2022a} or conventional (uniform) Bragg gratings \cite{Koufidis2022b}, we seek an ansatz to Eq.\ \eqref{Hill Equation}. As the ${\bf d}_{\left(\pm\right)}$-fields are axiomatically decoupled in a homogeneous bi-isotropic medium and, contrary to the spatial case of \cite{Jaggard1989, Flood1994, Flood1995}, strong coupling is expected to occur only between \emph{co-handed} counter-propagating modes, we elect 
\begin{equation}\label{Ansatz of Displacement}
    {\bf y}_{\left(\pm\right)}^{\rm t}=\left[y_{\left(\pm\right)}^{\rm t}\right]^{+}e^{-i\omega_{0,\left(\pm\right)}t}+\left[y_{\left(\pm\right)}^{\rm t}\right]^{-}e^{i\omega_{0,\left(\pm\right)}t}\,.
\end{equation}
The ``$+$'' (respectively, ``$-$'') superscript sign in the amplitudes denote forward and backward temporal reflections. The gyrotropy-perturbed angular frequency is
\begin{equation}\label{omega pm}
\omega_{0,\left(\pm\right)}=\omega_0\left(1\mp\frac{\bar{g}_r}{\bar{n}_r}\right)\,, \  {\rm where} \ \left|\bar{g}_r\right|<\bar{n}_r\,,
\end{equation}
with $\omega_0$ being the central (i.e., design) angular frequency (cf.\ Eq.\ (4) of \cite{Mostafa2023} but with the opposite harmonic convention).

\par Of course, ``negative time'' is merely an algebraic trick, as time only travels in one direction: the positive. Nonetheless, Eq.\ \eqref{Ansatz of Displacement} becomes meaningful when we reinstate the $e^{ikz}$ terms previously omitted (cf.\ Eq.\ \eqref{Solution for the Spatial Part}). Indeed, when a wave, $A^{+} e^{i(kz-\omega t)}$, encounters a temporal interface where the refractive index abruptly changes, reflection occurs, with the reflected wave typically given by $A^{-} e^{-i(kz+\omega t)}$. Now, since momentum is conserved, and the minus sign in front of the reflected wave's wavenumber lacks meaning, following the convention of \cite{Auld1968}, we take the complex conjugate, $\left(A^{-}\right)^{\ast} e^{i(kz+\omega t)}$, noting that both expressions have the \emph{same} real part. By omitting the spatial term in the exponential, Eq.\ \eqref{Ansatz of Displacement} is now fully justified. Such a time reversal is intimately linked to negative frequencies (cf.\ $\omega t=(-\omega)(-t)$) and can be understood in terms of phase conjugation (see Fig.\ 4 of \cite{Shaltout2019}) or negative refraction \cite{Pendry2008}.

\par For $\bar{g}_r=0$, Eq.\ \eqref{Ansatz of Displacement} reduces to the ansatz of achiral media (cf.\ Eq.\ (4) of \cite{Koufidis2023b}), whereas $\left|\bar{g}_r\right|=\bar{n}_r^{-1}$ marks the entrance to the negative refraction due to chirality regime \cite{Pendry2004}, but in the temporal domain (see \cite{Lasri2023}).
Manifestly, the expansion of Eq.\ \eqref{Ansatz of Displacement} aims to use the eigenmodes of pure circular birefringence (cf.\ Eq.\ \eqref{Beltrami Fields}) as the basis, and treat chirality as a perturbation (see Appx.\ B of \cite{Koufidis2022a} and \cite{Koufidis2023a}). As we prove in Sec.\ \ref{Negative refraction due to giant chirality}, the notation in Eq.\ \eqref{Ansatz of Displacement} remains nominal, since in a negative refracting state, the direction of phase propagation and the temporal handedness of counter-propagating modes are \emph{interchanged}.

\par Thence, substituting Eq.\ \eqref{Ansatz of Displacement} into Eq.\ \eqref{Hill Equation}, we apply the slowly varying envelope approximation. Under this adiabatic approximation, the time-dependence of the amplitudes appearing in Eq.\ \eqref{Ansatz of Displacement} is considered sufficiently mild, i.e., $|{{\rm d}^2y_{\left(\pm\right)}^{\rm t}}/{{\rm d} t^2}|\ll 2\omega_0|{{\rm d} y_{\left(\pm\right)}^{\rm t}}/{{\rm d} t}|$, except at the temporal boundaries. As derived in Appx.\ \ref{Derivation of the coupled-wave equations}, phase-matching potentially synchronous terms yields two distinct coupled-wave systems:
\begin{equation}\label{CWE}
   \frac{{\rm d}{\bf A}_{\left(\pm\right)}}{{\rm d}t}={\bf M}_{\left(\pm\right)}\cdot{\bf A}_{\left(\pm\right)}\,,
\end{equation}
where ${\bf A}_{\left(\pm\right)}=\left(\begin{matrix}\left[y_{\left(\pm\right)}^t\right]^+&\left[y_{\left(\pm\right)}^t\right]^-\\\end{matrix}\right)^{\intercal}$, with ${}^\intercal$ indicating transpose, and the characteristic matrices read
\begin{equation}\label{Characteristic Matrix}
    {\bf M}_{\left(\pm\right)}= \left(\begin{matrix}i\bar{\omega}_{\left(\pm\right)}&i\chi_{\left(\pm\right)}e^{-i2\delta\omega_{0,\left(\pm\right)} t}\\-i\chi_{\left(\pm\right)}e^{i2\delta\omega_{0,\left(\pm\right)} t}&-i\bar{\omega}_{\left(\pm\right)}\\\end{matrix}\right)\,.
\end{equation}
In the matrices of Eq.\ \eqref{Characteristic Matrix}, we have
\begin{equation}\label{barred omega}
    \bar{\omega}_{\left(\pm\right)} = \frac{\omega_{0,\left(\pm\right)}^2-\left[\bar{\theta}\right]_{\left(\pm\right)}}{2\omega_{0,\left(\pm\right)}}\,,
\end{equation}
the detuning parameters are given by 
\begin{equation}\label{detuning}
\delta\omega_{0,\left(\pm\right)}=\omega_0\left(1\mp\frac{\bar{g}_r}{\bar{n}_r}\right)-\frac{\Omega}{2}\,,
\end{equation}
and the identified coupling coefficients by
    \begin{equation}\label{Coupling Coefficient}    \chi_{\left(\pm\right)}=-\frac{\left[\delta\theta\right]_{\left(\pm\right)}}{4\omega_{0, \left(\pm\right)}}\,.
\end{equation}

\par As we show in Appx.\ \ref{Solutions of the coupled-wave equations}, the coupled-wave equations of Eq.\ \eqref{CWE} have analytic solutions in closed forms, which, for $t_0=0$, may be expressed in a matrix notation as
\begin{equation}\label{CWE Solutions}
    {\bf A}_{\left(\pm\right)}\left(t\right)={\bf S}_{\left(\pm\right)}\left(t\right)\cdot{\bf A_{\left(\pm\right)}}\left(0\right)\,.
\end{equation}
The components of the transfer matrices are
\begin{equation*}
    {\bf S}_{\left(\pm\right)}=\left(\begin{matrix}e^{- i\delta\omega_{0,\left(\pm\right)} t}p^{+}_{\left(\pm\right)}&e^{- i\delta\omega_{0,\left(\pm\right)}t}q^{+}_{\left(\pm\right)}\\e^{ i\delta\omega_{0,\left(\pm\right)} t}q^{-}_{\left(\pm\right)}&e^{ i\delta\omega_{0,\left(\pm\right)} t}p^{-}_{\left(\pm\right)}\\\end{matrix}\right)\,,
\end{equation*}
where by setting
\begin{equation}\label{Detuning Parameters Accurate}
    \delta\tilde{\omega}_{\left(\pm\right)} = \delta\omega_{0,\left(\pm\right)}+\bar{\omega}_{\left(\pm\right)}\,,
\end{equation}
we can write
\begin{align}\label{Definition of p and q}
    p^{\pm}_{\left(\pm\right)}&=\cosh{\left(\Delta_{\left(\pm\right)}t\right)}\pm i\frac{ \delta\tilde{\omega}_{\left(\pm\right)} }{\Delta_{\left(\pm\right)}}\sinh{\left(\Delta_{\left(\pm\right)}t\right)}\,,\nonumber
\\ 
q^{\pm}_{\left(\pm\right)}&=\pm i\frac{\chi_{\left(\pm\right)}}{\Delta_{\left(\pm\right)}}\sinh{\left(\Delta_{\left(\pm\right)}t\right)}\,,
\end{align}
with $\Delta_{\left(\pm\right)}=\left[\chi_{\left(\pm\right)}^2- \delta\tilde{\omega}_{\left(\pm\right)}^2\right]^{1/2}$. 

\section{Dispersion characteristics}
\label{Dispersion characteristics}
\subsection{Parametric amplification}
\label{Parametric amplification}
\begin{figure}[!t]
\includegraphics[width=0.5\linewidth]{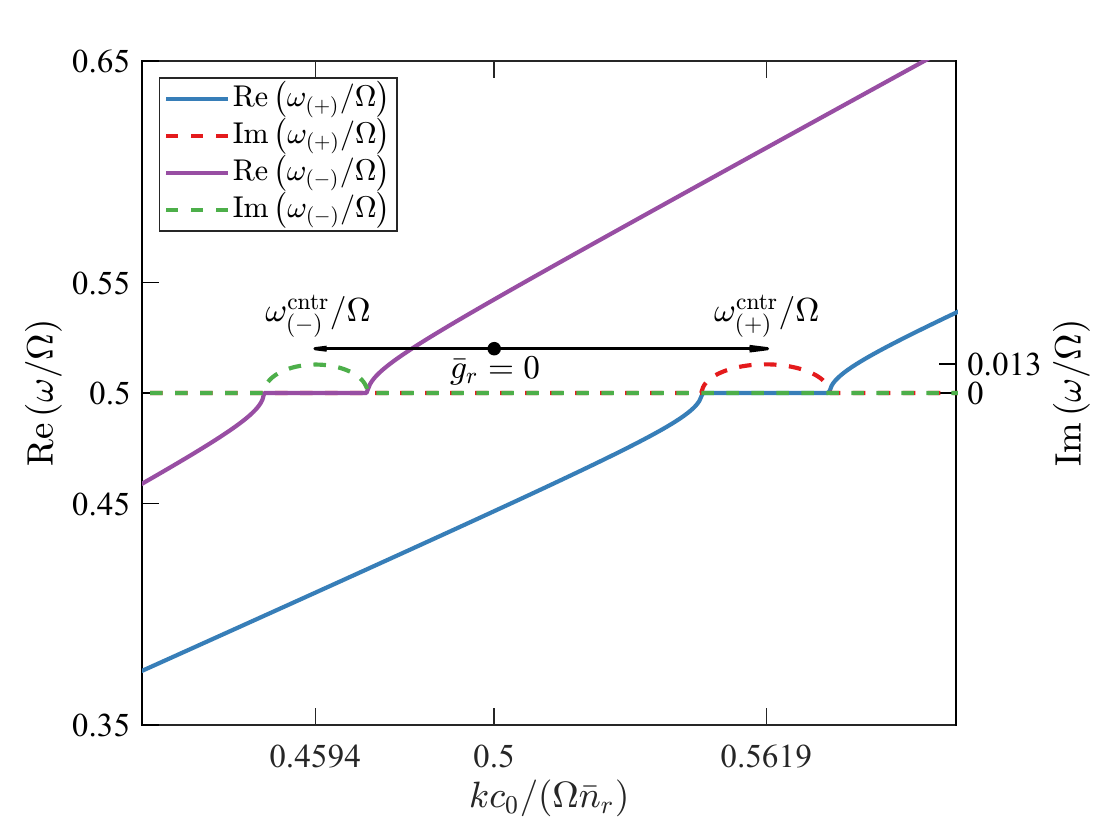}
\caption{Brillouin diagram illustrating the dispersion of each eigenmode supported by a periodically modulated bi-isotropic medium. When $\bar{g}_r=0$, the $\omega_{(\pm)}$ curves become degenerate, and the sole momentum gap is centered at 0.5. Switching the chirality on, the two branches of Eq.\ \eqref{dispersion relation} are separated, forming two distinct first-order momentum gaps, each for a different mode. For the left- (respectively, right-) handed mode, the resonances are shifted at the angular frequencies of Eq.\ \eqref{Center of Resonance LCP} (respectively, Eq.\ \eqref{Center of Resonance RCP}).  The base parameters are: for the permittivity, $\bar{\epsilon}_r=1$ and $\delta\epsilon_r=0.1$; for the permeability, $\bar{\mu}_r=1$ and $\delta\mu_r=0$; and for the chirality parameter, $\bar{g}_r=0.1$ and $\delta g_r=0.01$.
}
\label{Figure Brillouin Diagram}
\end{figure}

Developing the first row of Eq.\ \eqref{CWE Solutions} for the wave travelling forwards, it is immediately apparent from the various exponential terms that the possible angular frequencies are $\omega_{\left(\pm\right)} = {\Omega}/{2}\pm i\Delta_{\left(\pm\right)}$, or, equivalently,
\begin{equation}\label{dispersion relation}
\omega_{\left(\pm\right)} =  \frac{\Omega}{2} + s\left\{ \biggl[ \omega_0\biggl(1\mp\frac{\bar{g}_r}{\bar{n}_r}\biggr) +\bar{\omega}_{\left(\pm\right)}- \frac{\Omega}{2} \biggr]^2 - \chi_{\left(\pm\right)}^2 \right\}^{1/2}\,,
\end{equation}
where $s=\text{sign}(\delta\tilde{\omega}_{\left(\pm\right)})$; the choice of sign in front of the square root is dictated by the physical necessity that $\omega_{\left(\pm\right)}\rightarrow 0$ as $\omega_0\rightarrow 0$
and that $\omega_{\left(\pm\right)}\rightarrow\omega_{0,\left(\pm\right)}$ as $\omega_0\rightarrow +\infty$ (see \cite{Koufidis2022b, Koufidis2023b}). All branches of the dispersion relation in Eq.\ \eqref{dispersion relation} are plotted as a Brillouin diagram in Fig.\ \ref{Figure Brillouin Diagram}, where the $\omega_{\left(+\right)}$-curves are associated with left-handed eigenmodes, whereas the $\omega_{\left(-\right)}$-curves are associated with right-handed eigenmodes. Two first-order momentum gaps are formed, each corresponding to orthogonal modes. The modulation depth contributes to the elevation of the imaginary part in Eq.\ \eqref{dispersion relation}, thereby leading to the emergence of parametric amplification, as depicted by the dashed lines. The segments of the hyperbolas represent passbands, wherein, in the absence of absorption, a wave propagates without alteration.

\par For both polarizations, the nonzero imaginary part of the angular frequency inside the photonic bandgap gives rise to a regime wherein parametric amplification ensues \cite{Koutserimpas2018}. Such a behavior is expected regardless of the presence or absence of magneto-electric coupling. Indeed, a closer examination of the effective permittivity and permeability in Eq.\ \eqref{Effective Permittivity and Permeability} reveals that the impedance of the equivalent isotropic medium, $\eta_{\left(\pm\right)}=\left({\mu_{\left(\pm\right)}/{\epsilon_{\left(\pm\right)}}}\right)^{1/2}$, is independent of the chirality. Moreover, for $\bar{\epsilon}_r\neq\bar{\mu}_r$ (and/or $\delta\epsilon_r\neq\delta\mu_r$), the impedance varies over time, resulting in backscattering (i.e., bandgaps \cite{Pendry2021}), independently of $\bar{g}_r$ and $\delta g_r$. Nonetheless, as we subsequently discuss, chirality provides control over: (a) the locations of the momentum gaps, (b) their corresponding bandwidths, and (c) the amplification factors.

\par As anticipated, both the left- and the right-handed eigenmodes can be amplified, albeit at different angular frequencies. To identify the frequencies for which the largest moduli of the imaginary parts of $\omega$ are achieved, i.e., the centers of the two resonances, we can simply equate the detuning parameters of Eq.\ \eqref{detuning} to zero. Hence, the momentum gap for the left-handed eigenmodes will be centered at
\begin{subequations}\label{Center of Resonances}
    \begin{equation}\label{Center of Resonance LCP}
    \omega_{\left(+\right)}^{\rm cntr}=\frac{\Omega}{2}\left(1-\frac{\bar{g}_r}{\bar{n}_r}\right)^{-1}\,,
\end{equation}
whereas for the right-handed eigenmodes at
\begin{equation}\label{Center of Resonance RCP}
    \omega_{\left(-\right)}^{\rm cntr}=\frac{\Omega}{2}\left(1+\frac{\bar{g}_r}{\bar{n}_r}\right)^{-1}\,.
\end{equation}
\end{subequations}

\par Upon inspection of Eqs.\ \eqref{Center of Resonance LCP} and \eqref{Center of Resonance RCP}, it is clear that the presence of chirality shifts the centers of the resonances, towards the red or the blue, depending on the sense of rotation (i.e., the sign of $\bar{g}_r$). Importantly, such a shift is induced only by the presence of the chirality parameter's dc-term. Whether the chirality is time-modulated or not, $\bar{g}_r$ offers control over the location of the resonances, closely resembling the case where a structurally chiral medium is infiltrated by a chiral fluid \cite{Koufidis2022a}. We note that in the static medium of \cite{Koufidis2022a}, the shift induced by the presence of chirality is linear with respect to the Bragg wavelength. In our example, however, as is customary to examine the dispersion characteristics in the frequency domain, such dependence is manifested as an inverse law, but the shifting-mechanism is similar.

\par On the other hand, the modulation strength of the chirality affects the coupling between co-handed counter-propagating modes, which can be clearly demonstrated after some algebraic manipulations of Eq.\ \eqref{Coupling Coefficient}:
    \begin{equation}\label{Coupling Coefficient 2}    
    \chi_{\left(\pm\right)}=\left(\frac{\omega_0^2-\left[\bar{\theta}_2\right]_{\left(\pm\right)}}{2\omega_{0,\left(\pm\right)}}\right)\frac{\delta n_{r,\left(\pm\right)}}{\bar{n}_{r,\left(\pm\right)}}-\frac{\omega_0^2}{2\omega_{0,\left(\pm\right)}}\frac{\delta n_r}{\bar{n}_r}\,.
\end{equation}
As a validity check, in the absence of magneto-electric coupling, Eq.\ \eqref{Coupling Coefficient 2} reduces to $\chi_{\left(\pm\right)}=-\left({\omega_0}/{2}\right)\left({\delta n_r}/{\bar{n}_r}\right)$, thus corroborating Eq.\ (6) in \cite{Koufidis2023b}. The coupling coefficients are directly related to the momentum gap bandwidth, which can be estimated by setting $\Delta_{\left(\pm\right)}=0$. It turns out that the limiting frequencies are given by
    \begin{equation}\label{Bandwidth RCP and LCP}
    \frac{\Delta\omega_{\left(\pm\right)}}{\Omega}\approx\frac{\bar{n}_r\delta n_r}{2\bar{n}_{r,\left(\mp\right)}}
    \,,
\end{equation}
for each handedness, respectively.

\par
\begin{figure}[!t]
\includegraphics[width=0.5\linewidth]{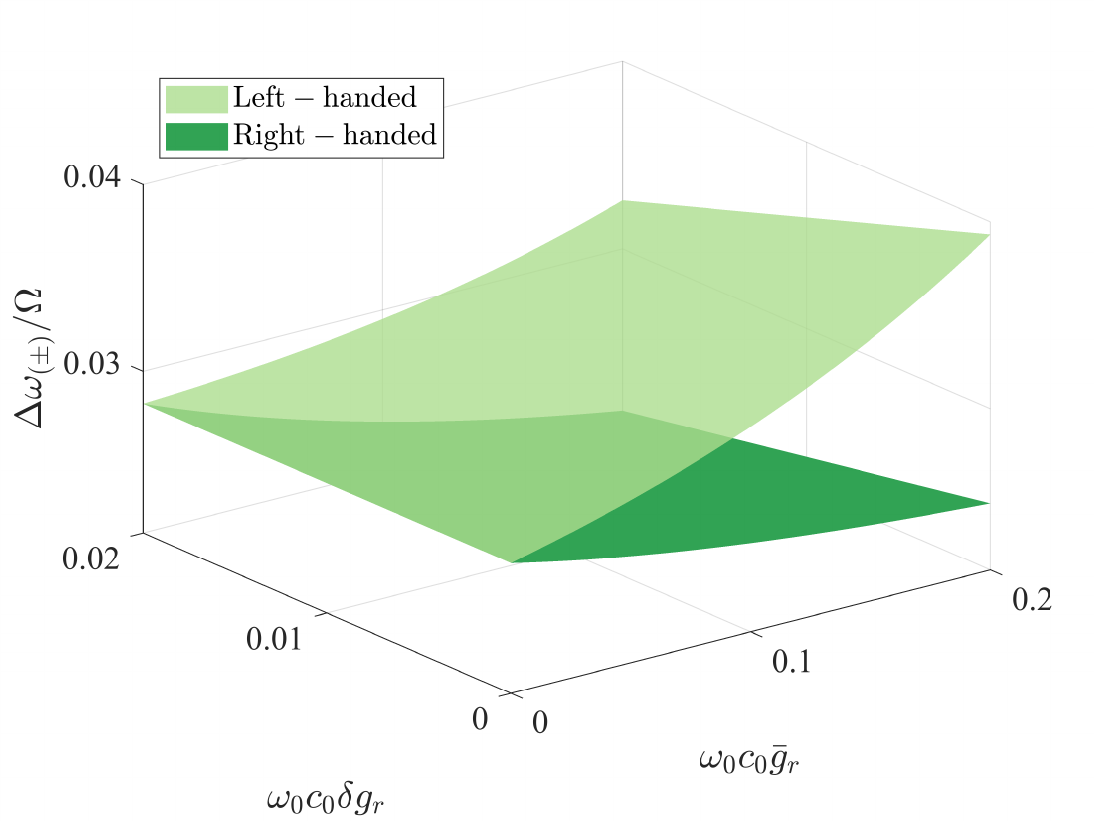}
\caption{The combined influence of the dc- and ac-terms of the chirality on the bandwidth of the momentum gap associated with each eigenmode for the parameters of Fig.\ \ref{Figure Brillouin Diagram}. The transparent (respectively, opaque) surface corresponds to the left- (respectively, right-) handed mode. For the left- (respectively, right-) handed mode, $\Delta\omega_{+}$ (respectively, $\Delta\omega_{-}$) increases (respectively, decreases) monotonically with $\bar{g}_r$ in a parabolic pattern, and there is a linear and consistently decreasing trend with $\delta g_r$ for both handednesses.
}
\label{Figure Bandwidths}
\end{figure}

 Although the simple formulae in Eq.\ \eqref{Bandwidth RCP and LCP} clearly illustrate the dependence of the bandwidths of the parametric amplification regimes on the dc-term of the chirality, they do not account for the influence of the chirality's modulation strength. Therefore, in Fig.\ \ref{Figure Bandwidths}, we present plots of the solutions to the equations $\Delta_{\left(\pm\right)}=0$ for various values of $\bar{g}_r$ and $\delta g_r$, without neglecting higher-order terms. As shown, for both the left- and the right-handed modes, the dependence of $\Delta\omega_{\left(+\right)}$ and $\Delta\omega_{\left(-\right)}$ on $\bar{g}_r$ follows a parabolic pattern, with the former increasing and the latter decreasing monotonically. Additionally, the dependence on $\delta g_r$ exhibits a linear and consistently decreasing trend for both handednesses.

\par Likewise, the presence of chirality also affects the maximum values of the imaginary parts of $\omega_{\left(\pm\right)}$, i.e., the amplification factors. To provide an estimation of this parameter, we shall examine the system on-resonance. In such case, Eq.\ \eqref{dispersion relation} yields
\begin{equation}\label{Amplification factor}
    \left|{\rm Im}\left(\omega_{\left(\pm\right)}\right)\right|_{\delta\tilde{\omega}_{\left(\pm\right)}=0}=\chi_{\left(\pm\right)}\,. 
\end{equation}
Hence, the dependence of the amplification factors on the chirality is clear (cf.\ the approximate expression of Eq.\ \eqref{Coupling Coefficient 2}); we revisit this point in Sec.\ \ref{Scattering coefficients}.

\subsection{Density of states}
\label{Density of States Sec}
\par Should we consider the number of angular frequencies lying between $\omega_0$ and $\omega_0+{\rm d}\omega_0$, where ${\rm d}\omega_0$ is infinitesimal, to be uniform, then the corresponding number of states in the range $\omega$ to $\omega+{\rm d}\omega$ is simply $\left({{\rm d}\omega}/{{\rm d}\omega_0}\right){\rm d}\omega_0$. Therefore, ${{\rm d}\omega}/{{\rm d}\omega_0}$ may be regarded as the ``density of states'' in angular frequency space. Differentiating Eq.\ \eqref{dispersion relation}, we arrive at the approximate expression
\begin{equation}\label{Density of States}
\frac{{\rm d}\omega_{\left(\pm\right)}}{{\rm d}\omega_0}\approx{\rm Re}\left[\frac{ \delta\tilde{\omega}_{\left(\pm\right)}\left(1\mp\bar{n}_r^{-1}\bar{g}_r\right)}{\left(\delta\tilde{\omega}_{\left(\pm\right)}^2-\chi_{\left(\pm\right)}^2\right)^{1/2}}\right]\,.
\end{equation}

\begin{figure}[!t]
\includegraphics[width=0.5\linewidth]{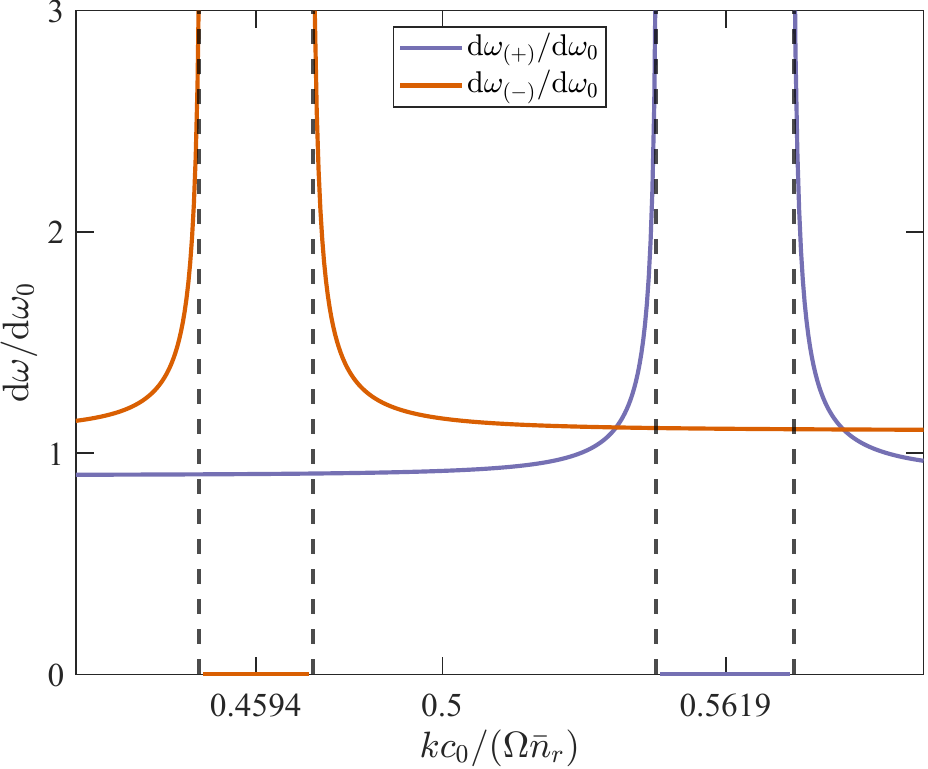}
\caption{Density of states in the vicinity of the first-order momentum gap for both eigenmodes, as per Eq.\ \eqref{Density of States} and for the parameters of Fig.\ \ref{Figure Brillouin Diagram}. By contrast to their spatial counterparts, in temporally modulated media, these curves are interpreted as the group velocity. At the edges of the momentum gaps (cf.\ the asymptotic dashed lines), the group velocity diverges. While not fundamentally forbidden, the group velocity becomes infinite at positions where standing waves are formed.
}
\label{Figure DensityStates}
\end{figure}

\par As illustrated in Fig.\ \ref{Figure DensityStates}, where Eq.\ \eqref{Density of States} is plotted, the density of states diverge at each band-edge of the momentum gaps (where $\Delta_{\left(\pm\right)}=0$). In the case under consideration, the density of states is the group velocity (aside from a factor of $c_0$) and not its inverse, as is the case in spatially modulated dielectric media. This raises the question: can the group velocity exceed the speed of light in vacuum? As extensively explained in Brillouin's book (see pp.\ 74--79 of \cite{Brillouin1960}), the group velocity can indeed surpass the speed of light at near-absorption frequencies (here, near amplification regimes). Nevertheless, the more physically pertinent \emph{signal} velocity is neither negative nor faster than the speed of light \cite{Noginov2009}. By contrast to stationary media, where infinite group velocity plays a pivotal role, e.g., in distributed feedback lasers \cite{Topf2014}, in temporal media, the group velocity diverges at positions where standing waves are formed \cite{Zurita2010}.

\section{Wave propagation through a finite ``time-slab''}
\label{Propagation through a finite time-slab}
\subsection{Scattering coefficients}
\label{Scattering coefficients}
We may now consider the electromagnetic response of a finite ``time-slab'' of the medium under discussion. Such a slab is defined as a homogeneous, isotropic, and achiral medium in which, at a certain moment, say, $t_{0}=0$, its permittivity, permeability, and chirality begin to vary periodically with time, as per Eq.\ \eqref{modulation profile}, and continue until a moment $t$ when the medium's parameters return to their pre-modulation values (cf.\ Fig.\ 1 of \cite{Ramaccia2020}). 

\par Hence, $\Delta t = t - t_{0}$ can be regarded as the duration of the slab, akin to the length of a Bragg grating. Modulating a medium for numerous periods (e.g., setting $\Delta t=10T$) undoubtedly presents significant experimental challenges, with notable concerns being raised regarding temporal dispersion and input power demands (see \cite{Hayran2022}). Nevertheless, recent experimental demonstrations involving highly nonlinear materials, such as ITO \cite{Alam2016}, in the epsilon-near-zero regime \cite{Vezzoli2018, Bohn2020, Bruno2020, Lustig2023}, hold substantial promise; for a comprehensive discussion on various experimental approaches we recommend Sec.\ V of \cite{Galiffi2022photonics}.

\par Considering the solutions of the coupled-wave equations given by Eq.\ \eqref{CWE Solutions} in terms of the transformed fields of Eq.\ \eqref{Field Transformation}, and by applying the only physically meaningful initial condition, $\left[d^{\rm t}_{\left(\pm\right)}\right]^{-}\left(t_0\right)=0$, as seen in \cite{Ramaccia2021}, we can define the backward and forward reflection coefficients in terms of the original excitation fields as
\begin{subequations}\label{reflection and transmission coefficients}
    \begin{align}
     r_{\left(\pm\right)}=&\frac{\left[d_{\left(\pm\right)}^{\rm t}\right]^{-}\left(t\right)}{\left[d_{\left(\pm\right)}^{\rm t}\right]^{+}\left(t_0\right)}=e^{\phi_{\left(\pm\right)}^{+}}q^{-}_{\left(\pm\right)}\,, \label{reflection coefficient}
     \\
     t_{\left(\pm\right)}=&\frac{\left[d_{\left(\pm\right)}^{\rm t}\right]^{+}\left(t\right)}{\left[d_{\left(\pm\right)}^{\rm t}\right]^{+}\left(t_0\right)}=e^{\phi_{\left(\pm\right)}^{-}}p^{+}_{\left(\pm\right)}\,,
     \label{transmission coefficient}
\end{align}
\end{subequations}
respectively, where 
\begin{equation}\label{Phase Term}
    \phi_{\left(\pm\right)}^{\pm}= -{\frac{1}{2}\int_{t_{0}}^t {{{\left[ {{\theta_1}} \right]}_{\left(\pm\right)}}{\rm d}t'} }\pm{ i\delta\omega_{0,\left(\pm\right)} t}\,.
\end{equation}

\par
\begin{figure}[ht]
\centering
\includegraphics[width=0.5\linewidth]{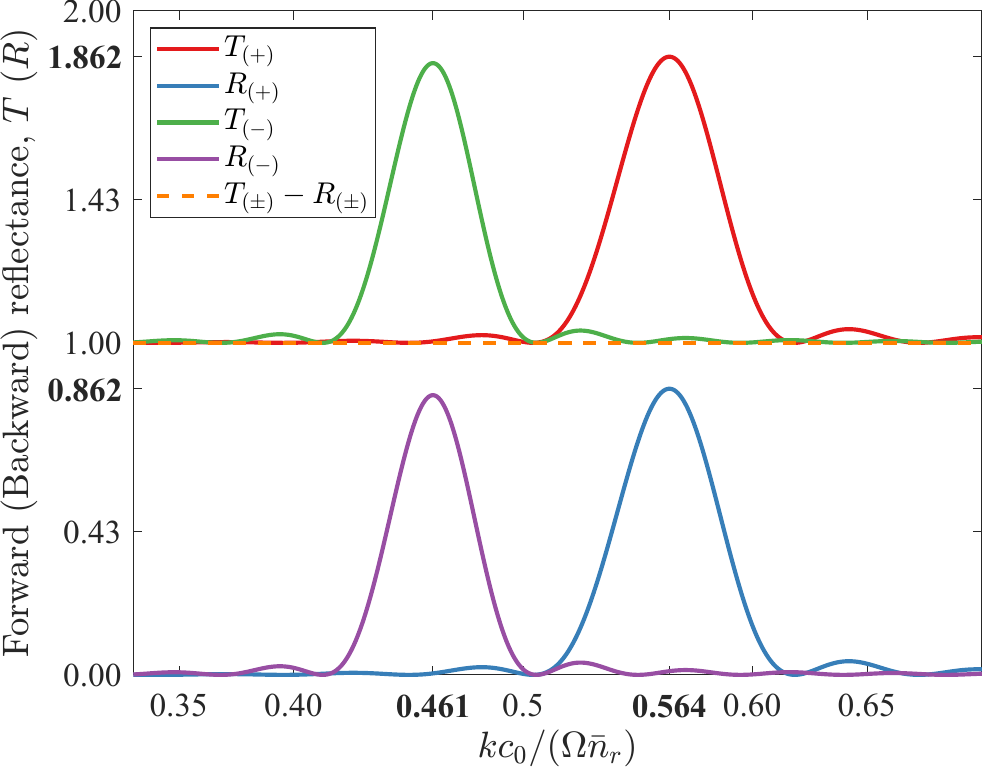}
\caption{Electromagnetic response of a finite ``time-slab'' of the considered medium for the parameters of Fig.\ \ref{Figure Brillouin Diagram} and for the slab's time duration being $\Delta t = 10T$. The resonance on the right- (respectively, left-) hand side of the figure corresponds to the parametric amplification of the left- (respectively, right-) handed eigenmode, inside the associated momentum gaps. The emphasized bold frequencies are those for which the maximum reflectance is achieved and are slightly different from those seen in Fig.\ \ref{Figure Brillouin Diagram}. An increment in the slab's time duration will force the resonances to become sharper, thus bringing the peak reflectances at the exact Bragg frequencies of Eqs.\ \eqref{Center of Resonances}. For both handednesses, the energy ``pseudo-conservation'' relation $T_{(\pm)} - R_{(\pm)} = 1$ is upheld (cf.\ the orange dashed line).}
\label{Figure OpticalResponse}
\end{figure}

 The formulae in Eq.\ \eqref{reflection and transmission coefficients} can be interpreted as the sum of the distributed (in time) chirality-preserving local reflections. However, these coefficients represent the total backward and  forward reflection only when dealing with impedance-matched media, where the ``surrounding'' medium is the pre- (and post-) modulation medium. However, we can match the auxiliary electric excitation fields, ${\bf d}_{\left(\pm\right)}$, and the auxiliary magnetic fields, ${\bf b}_{\left(\pm\right)}$, at the two temporal boundaries \cite{Morgenthaler1958, Xiao2014}, and derive the temporal Fresnel coefficients. These  turn out to be \cite{Caloz2020}
\begin{equation}\label{Temporal Fresnel Coefficients}
   \rho_{1\rightarrow2}=\frac{\eta_2-\eta_1}{2\eta_2} \ {\rm and} \  \tau_{1\rightarrow2}=\frac{\eta_2+\eta_1}{2\eta_2} \,,
\end{equation}
where $1\rightarrow2$ indicates the backward/forward reflection coefficient upon incidence from a medium with impedance $\eta_1$ to a medium with impedance $\eta_2$. Subsequently, by inserting the coefficients of Eq.\ \eqref{Temporal Fresnel Coefficients} into a two-by-two matrix, as done, e.g., in \cite{Ramaccia2021, Yang2023}, we can cascade the two matrices at the interfaces with the one appearing in Eq.\ \eqref{CWE Solutions}, when transformed back to the original excitation fields, to obtain the total reflection coefficients. The associated backward and forward intensity reflectances will then be $R_{\left(\pm\right)}=\left|\rho_{2\rightarrow 1}r_{\left(\pm\right)}\rho_{1\rightarrow 2}\right|^2$ and $T_{\left(\pm\right)}=\left|\tau_{2\rightarrow 1}t_{\left(\pm\right)}\tau_{1\rightarrow 2}\right|^2$, respectively. These are plotted for both polarizations in Fig.\ \ref{Figure OpticalResponse}, where the time duration of the slab is taken $\Delta t = 10T$ for illustration purposes, and the reflectances are plotted as a function of the spatial frequency at a snapshot at $t=\Delta t$.

\par The angular frequencies to which the resonance maxima correspond differ slightly from those shown in Figs.\ \ref{Figure Brillouin Diagram} and \ref{Figure DensityStates}. However, this discrepancy has a mundane origin, resulting from the always-in-phase waves not having enough time to produce a sharper, albeit crucially not steeper, resonance. Nevertheless, an increase in the permittivity's modulation depth augments the linewidth of the bandgap, further substantiating the Bragg-like features in the occurring amplification mechanism.

\par In Fig.\ \ref{Figure OpticalResponse}, we further plot the differences $T_{\left(\pm\right)}-R_{\left(\pm\right)}$, which are clearly equal to unity for all frequencies and both polarizations. This ``pseudo-conservation'' relation is a well-known property of periodically modulated temporal media \cite{Galiffi2022photonics}, and it resemblances the conservation relation of $\mathcal{PT}$-symmetric systems (see \cite{Ge2012}). Furthermore, it holds independently of the modulation profile, as indicated in \cite{Galiffi2022b}. It appears that such a conservation relation is also satisfied in the presence of a periodically modulated magneto-electric coupling. This is a predictable outcome, as for all the medium's parameters being real, the characteristic matrix of Eq.\ \eqref{Characteristic Matrix Transformed} is traceless and Hermitian, which guarantees that the transfer matrix of Eq.\ \eqref{CWE Solutions Transformed} is ${\rm SU}(1,1)$-symmetric \cite{Koufidis2022b}.

\par
\begin{figure}[ht]
\centering
\includegraphics[width=0.5\linewidth]{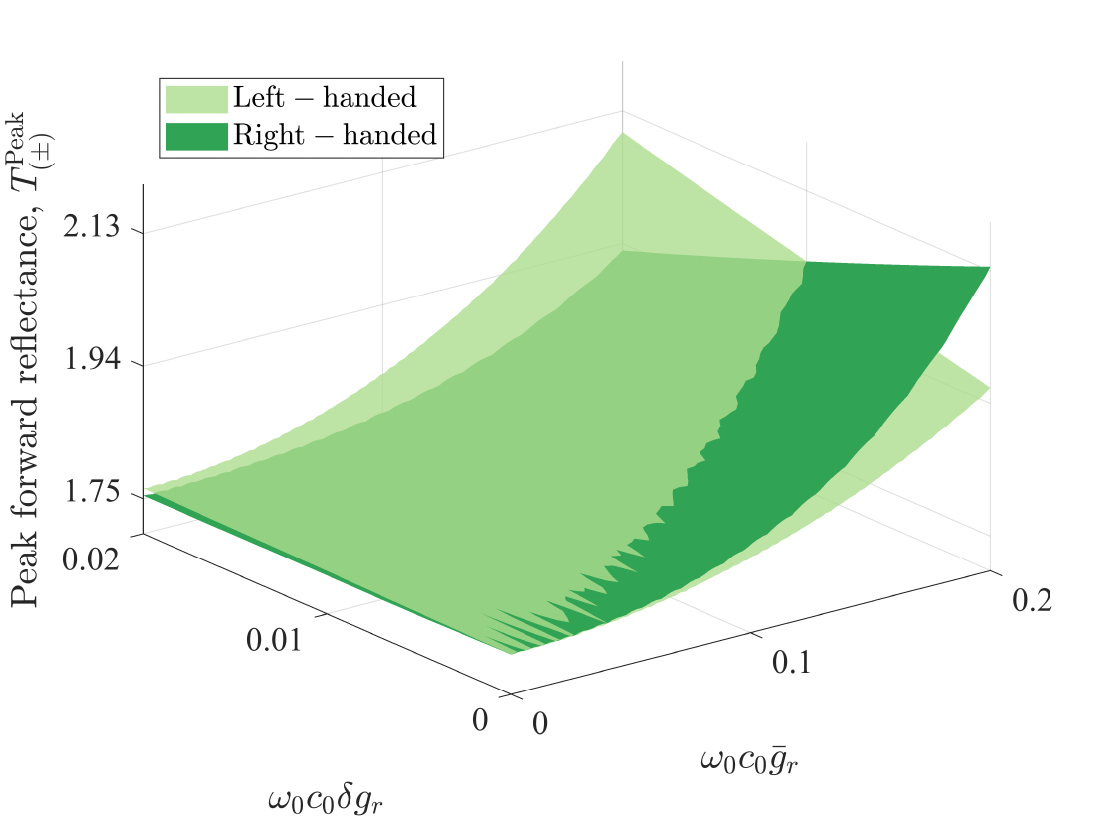}
\caption{The peak forward intensity reflectances for both handednesses, as per Eq.\ \eqref{On resonance transmittance}, for varying values of the dc- and ac-terms of the chirality within the context of Fig.\ \ref{Figure Brillouin Diagram} and for $\Delta t=10T$. Each point on the transparent (respectively, opaque) surface represents the maximum amplification achieved at the angular frequency of Eq.\ \eqref{Center of Resonance LCP} (respectively, Eq.\ \eqref{Center of Resonance RCP}) for the left- (respectively, right-) handed eigenmode. For both handednesses, the peak of the forward intensity reflectance increases with higher values of $\bar{g}_r$. With increasing values of $\delta g_r$, the peak for the left- (respectively, right-) handed eigenmodes increases (respectively, decreases).
}
\label{Figure Peak Transmittances}
\end{figure}

 As indicated by Eq.\ \eqref{Amplification factor}, the presence of magneto-electric coupling has a straightforward effect on the amplification factor. By quantitatively describing the enacted amplification process via, say, the forward intensity reflectance (i.e., the transmittance), we can promptly demonstrate the dependence of the maximum amplification on the dc- and ac-terms of the chirality. Focusing on the on-resonance case, Eqs.\ \eqref{Definition of p and q} and \eqref{transmission coefficient} dictate that
\begin{equation}\label{On resonance transmittance}
    T_{\left(\pm\right)}^{\rm Peak}=\left.\left|t_{\left(\pm\right)}\right|^2\right|_{\delta\omega_{0,\left(\pm\right)}=0}=\cosh^2\left(\chi_{\left(\pm\right)}\Delta t\right)\,,
\end{equation}
where for the considered geometry the $\tau$-coefficient of Eq.\ \eqref{Temporal Fresnel Coefficients} has been canceled out. From Eq.\ \eqref{On resonance transmittance} it is easy to see that the aforementioned conservation relation follows from the usual hyperbolic trigonometric identity. If the momentum gap is centered at the angular frequency given by Eq.\ \eqref{Center of Resonance LCP} (respectively, Eq.\ \eqref{Center of Resonance RCP}) for the left- (respectively, right-) handed eigenmode, the transparent (respectively, the opaque) surface in Fig.\ \ref{Figure Peak Transmittances} shows the maximum forward intensity reflectance achieved. Clearly, for both handednesses, the higher the value of $\bar{g}_r$, the greater the peak of the forward intensity reflectance. By contrast, with increasing values of $\delta g_r$, the peak of the forward intensity reflectance for the left- (respectively, right-) handed eigenmodes increases (respectively, decreases). 

\subsection{Negative refraction due to giant chirality}
\label{Negative refraction due to giant chirality}

\par Scrutinizing Eqs.\ \eqref{Ansatz of Displacement} and \eqref{omega pm}, it becomes apparent that for extreme values of the chirality parameter  \cite{Zhu2018, Wu2023, Koufidis2023c}, the signs of $\omega_{0,\left(\pm\right)}$ are swapped. Reminiscent of the static case, for $\bar{g}_r > \bar{n}_r$ (respectively, $\bar{g}_r < -\bar{n}_r$), Eq.\ \eqref{omega pm} implies that for the left- (respectively, right-) handed eigenmode, the direction of phase propagation between the (nominally) forward and backward propagating waves is interchanged. Notwithstanding, unlike the static scenario, the discussion on the modes' handedness is slightly more complicated.

\par Indeed, returning back to the Beltrami fields in Eq.\ \eqref{Beltrami Fields}, for $\bar{g}\in\left(-\bar{n}_r,\bar{n}_r\right)$, the eigenmodes propagating in the positive direction are
\begin{subequations}
    \begin{equation}\label{LCP eigenmodes}
        {{\bf E}^{+}_{\left(\pm\right)}}={Q_{\left(\pm\right)}^+}\frac{e^{i\left(kz- \omega_{0,\left(\pm\right)}t\right)}}{\sqrt{2}}\left(\begin{matrix}1\\\pm i\\\end{matrix}\right)\,,
    \end{equation}
    whereas those propagating in the negative direction are
    \begin{equation}\label{RCP eigenmodes}
        {\bf E}^{-}_{\left(\pm\right)}={Q_{\left(\pm\right)}^-}\frac{e^{i\left(kz+\omega_{0,\left(\pm\right)}t\right)}}{\sqrt{2}}\left(\begin{matrix}1\\\pm i\\\end{matrix}\right)\,,
    \end{equation}
\end{subequations}
where $Q_{\left(\pm\right)}^{\pm}$ are time-dependent amplitudes. We note that hitherto, the subscript notation ``$\pm$,'' indicating the polarization state of each wavefield, was associated with the spatial part of the field, and therefore signified the fields' sense of rotation in \emph{space}.

\par Should we drop the $kz$-term, by setting $Q^{+}_{\left(-\right)}=0$ and $Q^{+}_{\left(+\right)}=0$, respectively, Eq.\ \eqref{LCP eigenmodes} yields
\begin{subequations}
    \begin{equation}
    {\rm  Re}
   \left({\bf E}_{\left(\pm\right)}^{+}\right)=
   \frac{|Q_{\left(\pm\right)}^{+}|}{\sqrt2}
  \left(\begin{matrix}\cos{\left[\omega_{0,\left(\pm\right)}t+{  \rm arg}\left(Q_{\left(\pm\right)}^{+}\right)\right]}\\ \pm\sin{\left[\omega_{0,\left(\pm\right)}t+{  \rm arg}\left(Q_{\left(\pm\right)}^{+}\right)\right]}\\\end{matrix}\right)\,,
    \label{LH Electric Field plus}
\end{equation}
provided that ${\rm Im}\left(\bar{n}_r\right)=0$; otherwise, an exponential damping in the amplitudes must be included. Accordingly, upon setting $Q^{-}_{\left(-\right)}=0$ and $Q^{-}_{\left(+\right)}=0$ in Eq.\ \eqref{RCP eigenmodes}, we obtain, respectively,
    \begin{equation}
    {\rm  Re}
   \left({\bf E}_{\left(\pm\right)}^{-}\right)=
   \frac{|Q_{\left(\pm\right)}^{-}|}{\sqrt2}
  \left(\begin{matrix}\cos{\left[\omega_{0,\left(\pm\right)}t+{  \rm arg}\left(Q_{\left(\pm\right)}^{-}\right)\right]}\\ \mp\sin{\left[\omega_{0,\left(\pm\right)}t+{  \rm arg}\left(Q_{\left(\pm\right)}^{-}\right)\right]}\\\end{matrix}\right)\,.
    \label{RH Electric Field plus}
\end{equation}
\end{subequations}

\par For $\bar{g}_r \in (-\bar{n}_r, \bar{n}_r)$, the interpretation of Eqs.\ \eqref{LCP eigenmodes} and \eqref{RCP eigenmodes} is straightforward: they inform us about which direction of phase propagation is considered positive. Regarding the associated handedness of each mode, we underline that within the context of purely temporal media, one must distinguish between spatial and temporal handedness. The former, found on examining Eqs.\ \eqref{LCP eigenmodes} and \eqref{RCP eigenmodes} by taking a snapshot of the field, can be a right- or a left-handed helix \emph{in space}. The latter, on the other hand, found on examining Eqs.\ \eqref{LH Electric Field plus} and \eqref{RH Electric Field plus} for a fixed point in space as time evolves, can be a right- or a left-handed spiral \emph{in time}. In our problem, the spatial helix is fixed due to conservation of momentum, so the decomposed wavefields remain decoupled. However, the temporal spiral may well change handedness.

\par In particular, for $\bar{g}_r>\bar{n}_r$ (respectively, $\bar{g}_r<-\bar{n}_r$), $\omega_{0,\left(+\right)}$ (respectively, $\omega_{0,\left(-\right)}$) becomes negative. As a result, the direction of phase propagation between Eq.\ \eqref{LCP eigenmodes} and Eq.\ \eqref{RCP eigenmodes} for the ``$+$'' (respectively, ``$-$'') mode is changed, and the same thing happens for the sense of the temporal rotation, as the $y$-component of the vectors in Eqs.\ \eqref{LH Electric Field plus} and \eqref{RH Electric Field plus} changes signs. Under corresponding preconditions,  giant chirality can grant access to higher order Bragg resonances (n.b.\ Hill’s determinant of the Floquet theory predicts momentum bandgaps centered at $\omega=N{\Omega}/{2}$, $N\in\mathbb{N}$ \cite{Koutserimpas2018b}). Naturally, coupled-wave theory concentrated around the first harmonic needs to be modified to that of Sec.\ 4 in \cite{Koutserimpas2022}, but the intuition is clear.

\subsection{Polarization dynamics}
\label{Polarization Dynamics}
No definitive discussion on bi-isotropic media can be reached without exploring the influence of chirality on the polarization of electromagnetic waves when transmitted through such media. However, in the context of pure temporal modulation, the conventional interpretation of polarization can be somewhat misleading. Indeed, consider the transverse component of the electric field of a plane electromagnetic wave propagating along, say, the $z$-axis, ${\bf E}_\bot=\left(E_x,E_y\right)^{\intercal}e^{i\left(kz-\omega t\right)}$. It is possible to fix $z=z_0$ in the laboratory and track the loci of the trace of the electric field vector as time progresses (for harmonic waves, $\omega t\in\left[0,2\pi\right]$ is sufficient). Clearly, this approach is not directly applicable in a temporal slab, given the changing nature of $\omega$.

\par
\begin{figure}[ht]
\centering
\includegraphics[width=0.5\linewidth]{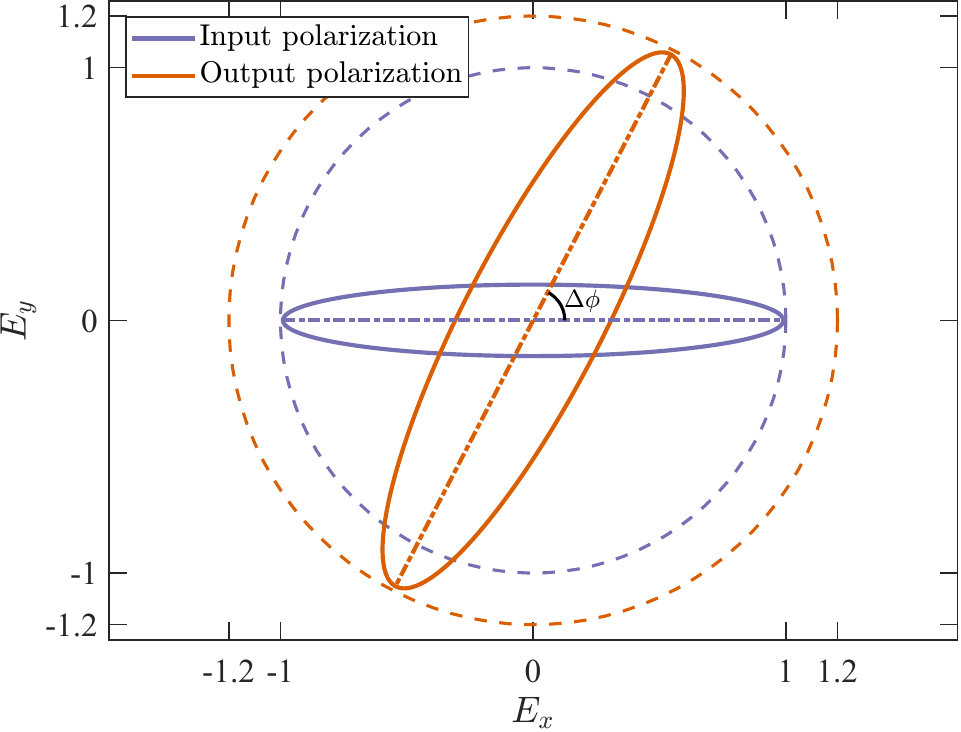}
\caption{Elliptical trajectory of the electric field just before (purple ellipse) and right after (orange ellipse) the passage of the wave through a temporally modulated bi-isotropic medium. The parameters are those of Fig.\ \ref{Figure Brillouin Diagram}, and the snapshot has been taken at $t_0=2.5T$.Both ellipses are inscribed in circles with diameters equal to the corresponding major axes of the ellipses, so that amplification is demonstrated as the area difference between the two circles. For such a time duration, the polarization rotation due to temporal circular birefringence is $\Delta\phi\approx {\pi}/{3}$.
}
\label{Figure Polarization Rotation 2D}
\end{figure}

 Nevertheless, it is feasible to parameterize the field in terms of the spatial component $z$ instead of $t$ and observe the loci of the tip of the electric field vector at a specific moment over a cycle of points along the $z$-axis (i.e., for $kz\in\left[0,2\pi\right]$). The objective of such a parameterization is clear: to convey how the electric field vector is influenced by propagation through a medium. This is illustrated in Fig.\ \ref{Figure Polarization Rotation 2D}, where the pre-modulation elliptically polarized wave is depicted alongside the post-modulation wave. The transmitted wave is evidently still elliptically polarized, albeit with the plane of polarization rotated. Furthermore, since the incident wave has been deliberately chosen to be elliptically polarized, containing unequal amounts of right and left circular polarizations, upon entering the temporal slab, its angular frequency undergoes a split due to temporal circular birefringence. Subsequently, as the angular frequency of each polarization falls within distinct momentum gaps, each component undergoes amplification, albeit in a distinct manner due to the different amplification factors for each handedness. The recombined wave at the output is then amplified, and this characteristic is manifested as the area difference between the two circles in Fig.\ \ref{Figure Polarization Rotation 2D}, each with a diameter equal to the major axis of the respective ellipse.

\par Another merit of the depiction in Fig.\ \ref{Figure Polarization Rotation 2D} is that it allows observations on the impact of the simultaneous existence of modulated chirality and permittivity/permeability on polarization. In fact, as can be seen, even for small values of chirality where the scattered intensities reach levels similar to those of the achiral case (e.g., for the parameters of Fig.\ \ref{Figure Brillouin Diagram}, ${T_{\left(+\right)}^{\rm chiral}}/{T_{\left(+\right)}^{\rm achiral}}\approx 1.055$), the polarization rotation is significant (in this instance, $\Delta\phi\approx {\pi}/{6}$). Furthermore, when the permittivity is modulated in such a fashion that parametric amplification occurs, we observe that the major (and minor) axes of the initial ellipse are increased, signifying an amplified wave.

\par With the aforementioned considerations, it is arguably more illustrative to refer to ``polarization dynamics,'' comprehended in the traditional sense of depicting how the tip of the electric field vector evolves over time. Thus, for \(t \in (-\infty, 0^{-}]\), let us consider only the forward traveling plane electromagnetic wave 
\begin{equation}\label{Input Polarization}
    \mathbf{E}_\bot^{\rm in} = \alpha_{\left(+\right)}e^{i\left(kz-\omega_0 t\right)} \hat{\mathbf{e}}_{(+)}+ \alpha_{\left(-\right)}e^{i\left(kz-\omega_0 t\right)} \hat{\mathbf{e}}_{(-)}\,,
\end{equation}
where $\alpha_{(+)}$ and $\alpha_{(-)}$ denote the amounts of left and right circular polarizations, respectively; for $\alpha_{(+)}\neq\alpha_{(-)}$, Eq,\ \eqref{Input Polarization} describes an elliptically polarized wave. Upon entering the temporal slab at \(t=0^+\), and for \(t \in (0^+, \Delta t^-]\), the angular frequency undergoes a downshift (respectively, upshift) for the left- (respectively, right-) handed mode. Therefore, the two circular modes of Eq.\ \eqref{Input Polarization} experience different frequencies, a manifestation of temporal circular birefringence (cf.\ Fig.\ 2 in \cite{Yin2022}). Then, for \(t \in [\Delta t^+, \infty)\), the post-modulation electric field is
\begin{align*}\label{Output Polarization}
    \mathbf{E}_\bot^{\rm out} &= \left(\tau_{2\rightarrow1}t_{\left(+\right)}e^{-i\omega_{0,\left(+\right)}\Delta t} \tau_{1\rightarrow2}\right)\alpha_{\left(+\right)}e^{i\left(kz-i\omega_0 t\right)} \hat{\mathbf{e}}_{(+)} \nonumber \\
   &+\left(\tau_{2\rightarrow1}t_{\left(-\right)}e^{-i\omega_{0,\left(-\right)}\Delta t} \tau_{1\rightarrow2}\right)\alpha_{\left(-\right)}e^{i\left(kz-i\omega_0 t\right)} \hat{\mathbf{e}}_{(-)}\,,
\end{align*}
whereby it is evident that by the time the wave exits the temporal slab, the plane of its elliptical polarization has been approximately rotated by \(\omega_0(\bar{g}_r/\bar{n}_r)\Delta t\), which is the temporal analog of the result obtained in Sec.\ III of \cite{Jaggard1989}.

\par
\begin{figure}[ht]
\centering
\includegraphics[width=0.5\linewidth]{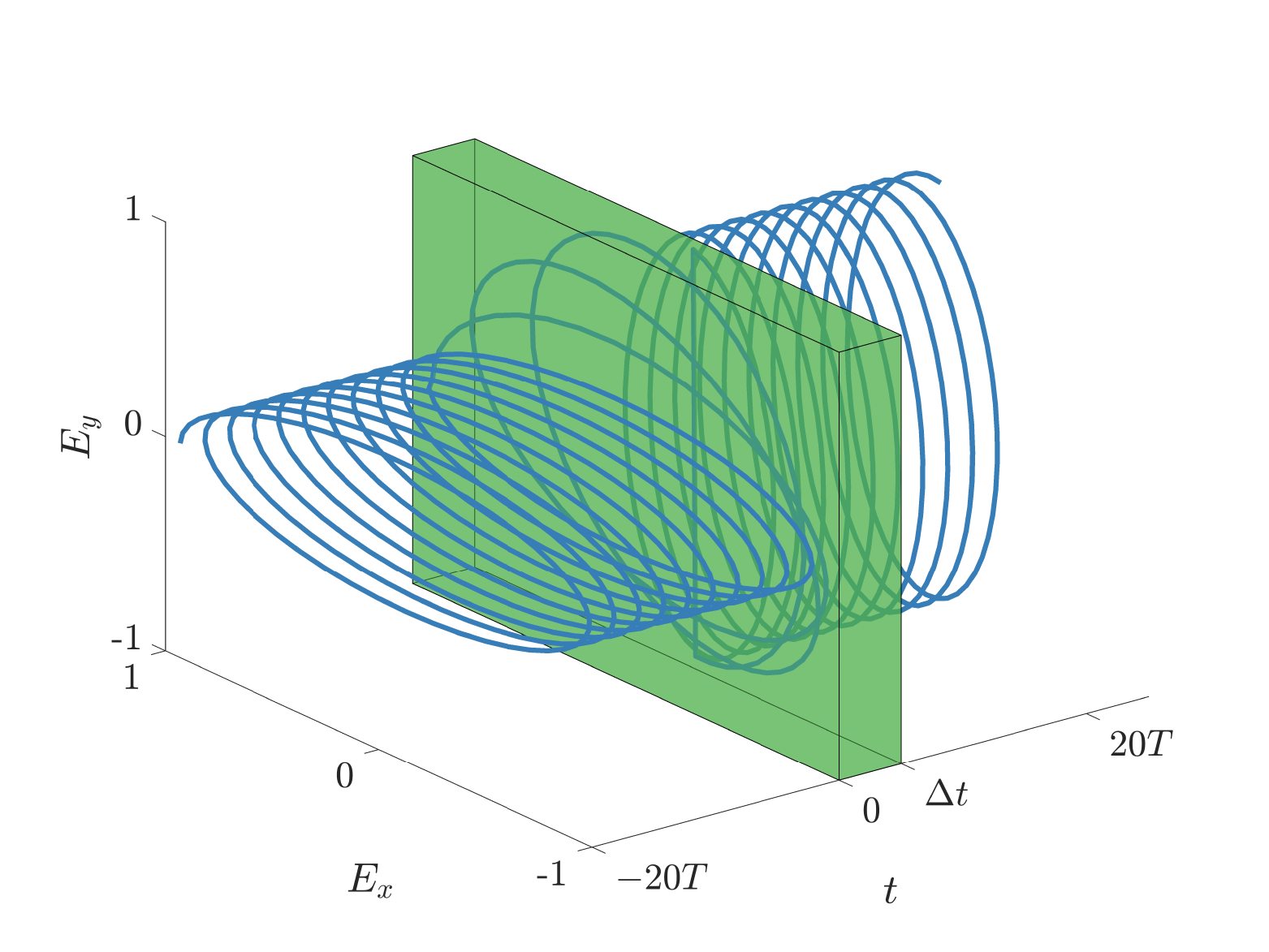}
\caption{Dynamic evolution of the electric field vector in time as an initially elliptically polarized light, composed of circular components $\alpha_{(+)}=0.4$ and $\alpha_{(-)}=\left(1-\alpha_{(+)}^2\right)^{1/2}$, passes through a temporal slab of a time-varying bi-isotropic medium with parameters as those of Fig.\ \ref{Figure Brillouin Diagram}. The slab's time duration has been set to $\Delta t=\left({\bar{n}_r}/{\bar{g}_r}\right)\left({\Delta\phi}/{\Omega}\right)=5T$, so that the initial plane of polarization is rotated, approximately, by ${\pi}/{2}$ upon transmission. Note the field discontinuity as the wave exits the temporal slab due to the failure of impedance matching, precisely at $t=\Delta t$.
}
\label{Figure Polarization Rotation 3D}
\end{figure}

 Consequently, similar to their spatial counterparts, temporally modulated bi-isotropic media can function as \emph{active} polarization rotators, with their optical rotatory power defined in terms of ${}^{\circ}/{\rm s}$ instead of ${}^{\circ}/{\rm mm}$. Moreover, owing to the dispersion properties of media with periodically time-varying permittivity, the temporal slab also acts as a gain medium, amplifying the output wave (cf.~the major and minor axes of the output ellipse compared to those of the input in Fig.\ \ref{Figure Polarization Rotation 2D} and in Fig.\ \ref{Figure Polarization Rotation 3D} for $t>\Delta t^{+}$). Remarkably, the proposed medium serves as an ideal platform for polarization control, where we can simultaneously control both the polarization rotation and the level of the output signal. Applications in, e.g., optical modulation \cite{Ma2023}, are expected to emerge.

\par If we wish to rotate the plane of polarization of an elliptically polarized wave by an angle $\Delta\phi$, then for a particular value of the dc-term of the chirality, we may set the time duration of the slab to $\Delta t=\left({\bar{n}_r}/{\bar{g}_r}\right)\left({\Delta\phi}/{\Omega}\right)$. In Fig.\ \ref{Figure Polarization Rotation 3D}, the dynamic evolution of the electric field with time is depicted as an elliptically polarized wave encounters a medium whose permittivity, permeability, and chirality periodically vary with time.
If the desired rotation is $\Delta\phi={\pi}/{2}$, by appropriately choosing the slab's time duration, the plane of an elliptically polarized wave is rotated by ${\pi}/{2}$. Such a choice of the slab's time duration does not exactly achieve the targeted $\Delta\phi$, as this would require considering the phase response of the optically active temporal Bragg grating. Nevertheless, the approximation is evidently sufficient.

\section{Conclusions}
\label{Conclusions}
The physics of dynamic media, where permittivity varies with time rather than space, is at the forefront of photonic research, coupled with experimental validation and demonstration. In this study, we harnessed the temporal dimension and combined it with chirality to investigate a temporally modulated bi-isotropic medium with constitutive parameters that are periodic functions of time. This medium was found to support two orthogonal circular polarizations with distinct frequencies. In the vicinity of these frequencies and under specific conditions found in our analysis, regimes of parametric amplification emerge. Notably, the presence of chirality results in the formation of two momentum gaps, each corresponding to orthogonal polarizations, whereby modes of different handedness exhibit amplification.

\par Utilizing coupled-wave theory, we elucidate how the presence of chirality influences the centers, bandwidths, and associated amplification factors of the first-order momentum gaps. The electromagnetic response of a ``time-slab'' of the medium validates our findings and further clarifies the energy ``pseudo-conservation'' relationship in such media. For weak chirality, the intensity scattering coefficients reach similar levels to the achiral case, whereas for chirality values comparable to the medium's time-averaged refractive index, we demonstrate the mechanism under which one of the two modes enters a negative refracting state.

\par Importantly, by presenting an alternative parameterization of polarization and examining its dynamical evolution, we show that even for weak values of chirality, where the scattering coefficients are at the same levels as in the achiral case, polarization rotation can become significant. With the simultaneous existence of periodicity in both permittivity and chirality, we outline the route towards the ideal optical modulator. Through this, we aim to simultaneously control the polarization of the signal along with its output power.

\par Potential applications of the proposed medium span diverse areas, ranging from enhancing electron transport in organic semiconductors \cite{Campbell2013} and advancing broadband wireless communications \cite{Hu2023a} to enabling vectorial polarimetry imaging \cite{Hu2023b}, optimizing efficiency in OLED displays \cite{Li2015}, and facilitating the development of asymmetrically transmitting photonic devices \cite{Han2021}.

\section*{Acknowledgments}
S.\ F.\ K.\ is financially supported by the Bodossaki Foundation. T.\ T.\ K.\ is supported by the Postdoc Mobility Fellowship from the Swiss National Science Foundation (SNSF), Grant No.\ 203176, and the ``Stamatis G. Mantzavinos'' Postdoctoral Scholarship from the Bodossaki Foundation. F. M. is supported  by the Air Force Office of Scientific Research with Grant No. FA9550-22-1-0204 through Dr. Arje Nachman.

\section*{Appendices}
\appendix 

\section{Fourier expansions of the characteristic coefficients}
\label{Fourier expansions of the characteristic coefficients}
Considering the modulation profile of each relative parameter, as given by Eq.\ \eqref{modulation profile}, the time-local refractive index is approximated as
\begin{equation*}\label{time local n}
    n_r  =\left(\epsilon_r\mu_r\right)^{1/2}\approx\bar{n}_r+\delta n_r\frac{e^{i\Omega t}+e^{-i\Omega t}}{2}\,,
\end{equation*}
where $\bar{n}_r=\left(\bar{\epsilon}_r\bar{\mu}_r\right)^{1/2}$ and $\delta n_r=\left({\bar{\eta}}/{2}\right)\delta\epsilon_r+\left(2\bar{\eta}\right)^{-1}\delta\mu_r$, with $\bar{\eta}=\left({\bar{\mu}_r}/{\bar{\epsilon}_r}\right)^{1/2}$. Accordingly, 
\begin{equation*}\label{time local n pm}
    n_{\left(\pm\right)}  =\left(\epsilon_{\left(\pm\right)}\mu_{\left(\pm\right)}\right)^{1/2}\approx\bar{n}_{r,\left(\pm\right)}+\delta n_{r,\left(\pm\right)}\frac{e^{i\Omega t}+e^{-i\Omega t}}{2}\,,
\end{equation*}
where $\bar{n}_{r,\left(\pm\right)}=\bar{n}_r\pm \omega c_0 \bar{g}_r$ and $\delta n_{r,\left(\pm\right)}=\delta n_r\pm\omega c_0\delta g_r$. 

\par As a prelude for approximating the $\left[\theta_1\right]_{\left(\pm\right)}$ coefficients, we find the Fourier expansion 
\begin{equation}\label{h function}
    h_{\left(\pm\right)}=1\pm\frac{g_r}{n_r}\approx\bar{h}_{\left(\pm\right)}+\delta h_{\left(\pm\right)}\frac{e^{i\Omega t}+e^{-i\Omega t}}{2}\,,
\end{equation}
where $\bar{h}_{\left(\pm\right)}={\bar{n}_{r,\left(\pm\right)}}/{\bar{n}_r}$ and $\delta h_{\left(\pm\right)}={\delta n_{r,\left(\pm\right)}}/{\bar{n}_r}-\left({\bar{n}_{r,\left(\pm\right)}}/{\bar{n}_r^2}\right)\delta n_r$. Furthermore, we calculate
\begin{equation*}\label{approximate f-relation}
    \frac{1}{f}\frac{\partial f}{\partial t}=i\Omega\frac{e^{i\Omega t}-e^{-i\Omega t}}{{2\bar{f}}/{\delta f}+e^{i\Omega t}+e^{-i\Omega t}}=-\Omega\frac{\sin{\Omega t}}{{\bar{f}}/{\delta f}+\cos{\Omega t}}\,.
\end{equation*}
Since $|\cos{\Omega t}|\leq 1$ and ${\bar{f}}/{\delta f}\gg 1$, the Taylor expansion of the denominator of the fraction above leads to
\begin{equation}\label{approximate f-relation approx}
    \frac{1}{f}\frac{\partial f}{\partial t}\approx i\Omega\frac{\delta f}{\bar{f}}\frac{ e^{i\Omega t}-e^{-i\Omega t}}{2}\,.
\end{equation}
\par Expanding now $\left[\theta_1\right]_{\left(\pm\right)}$ as
\begin{equation*}\label{theta 1 expansion}
     \left[\theta_1\right]_{\left(\pm\right)}  = \frac{1}{\mu_{\left(\pm\right)}}\frac{{\rm d} \mu_{\left(\pm\right)}}{{\rm d} t}=\frac{1}{\mu_r}\frac{{\rm d}\mu_r}{{\rm d}t}+\frac{1}{h_{\left(\pm\right)}}\frac{{\rm d}h_{\left(\pm\right)}}{{\rm d}t}\,,
\end{equation*}
we utilize Eqs.\ \eqref{h function} and \eqref{approximate f-relation approx}, and write
\begin{equation*}\label{Theta 1 approx}
    \left[\theta_{1}\right]_{\left(\pm\right)}  \approx \left[\theta_{m}\right]_{\left(\pm\right)}\frac{e^{i\Omega t}-e^{-i\Omega t}}{2}\,,
\end{equation*}
where $\left[\theta_{m}\right]_{\left(\pm\right)}=i\Omega\left({\delta \mu_r}/{\bar{\mu}_r}+{\delta n_{r,\left(\pm\right)}}/{\bar{n}_{r,\left(\pm\right)}}-{\delta n}/{\bar{n}_r}\right)$.

\par Regarding the $\left[\theta_2\right]_{\left(\pm\right)}$ coefficients, we can readily write
\begin{equation}\label{Theta 2 Appox}
    \left[\theta_2\right]_{\left(\pm\right)}  =\frac{k^2c_0^2}{n_{\left(\pm\right)}^2  }\approx\left[\bar{\theta}_2\right]_{\left(\pm\right)}+\left[\delta\theta_2\right]_{\left(\pm\right)}\frac{e^{i\Omega t}+e^{-i\Omega t}}{2}\,,
\end{equation}
where $\left[\bar{\theta}_2\right]_{\left(\pm\right)} ={k^2c_0^2}/{\bar{n}_{r,\left(\pm\right)}^2}$ and $\left[\delta\theta_2\right]_{\left(\pm\right)}=-2{k^2c_0^2\delta n_{r,\left(\pm\right)}}/{\bar{n}_{r,\left(\pm\right)}^3}$, provided that $\delta n_{r,\left(\pm\right)}\ll 1$. Conversely, we have avoided the formal integral-definitions for the $\left[\theta_{2}\right]_{\left(\pm\right)}$-coefficients, as seen in Appx.\ D of \cite{Zurita2009}, and contend that the approximations of Eq.\ \eqref{Theta 2 Appox} are indeed acceptable for weak modulation depths.

\section{Derivation of the coupled-wave equations}
\label{Derivation of the coupled-wave equations}
Calculating the first and second time-derivatives of the ansatz in Eq.\ \eqref{Ansatz of Displacement}, 
we can neglect second time-derivatives in the amplitudes, within the context of the slowly varying envelope approximation. Thence, substituting Eq.\ \eqref{Ansatz of Displacement} and the two calculated expressions into Eq.\ \eqref{Hill Equation}, also considering Eq.\ \eqref{dc and amplitude theta}, we obtain
\begin{multline}\label{Lengthy Equation}
-2i\omega_{0,\left(\pm\right)}\frac{{\rm d}[y_{\left(\pm\right)}^t]^+}{{\rm d}t}e^{-i\omega_{0,\left(\pm\right)}t}+2i\omega_{0,\left(\pm\right)}\frac{{\rm d}[y_{\left(\pm\right)}^t]^-}{{\rm d}t}e^{i\omega_{0,\left(\pm\right)}t}
+\left(\left[\bar{\theta}\right]_{\left(\pm\right)}-\omega_{0,\left(\pm\right)}^2\right)[y_{\left(\pm\right)}^t]^+e^{-i\omega_{0,\left(\pm\right)}t} \\
+\left(\left[\bar{\theta}\right]_{\left(\pm\right)}-\omega_{0,\left(\pm\right)}^2\right)[y_{\left(\pm\right)}^t]^-e^{i\omega_{0,\left(\pm\right)}t}
+[y_{\left(\pm\right)}^t]^+[\delta\theta]_{\left(\pm\right)}\frac{e^{i\left(\mathrm{\Omega}-\omega_{0,\left(\pm\right)}\right)t}+e^{-i\left(\mathrm{\Omega}+\omega_{0,\left(\pm\right)}\right)t}}{2}\\
+[y_{\left(\pm\right)}^t]^-[\delta\theta]_{\left(\pm\right)}\frac{e^{i\left(\mathrm{\Omega}+\omega_{0,\left(\pm\right)}\right)t}+e^{-i\left(\mathrm{\Omega}-\omega_{0,\left(\pm\right)}\right)t}}{2}=0\,.
\end{multline}
We note the subtlety that, in this instance, contrary to that of \cite{Koufidis2023b}, the terms $\left[\bar{\theta}\right]_{\left(\pm\right)}- \omega_{0,\left(\pm\right)}^2$ ought to be retained.

\par We now move on to applying the classic perturbation theory by introducing the phase-mismatch, $\delta\omega_{0,\left(\pm\right)}$, of Eq.\ \eqref{detuning}. Whence, the various exponential terms appearing in Eq.\ \eqref{Lengthy Equation} become:
\begin{align*}
e^{i\left(\mathrm{\Omega}-\omega_{0,\left(\pm\right)}\right)t} & =e^{i\left({\Omega}/{2}-\delta\omega_{0,\left(\pm\right)}\right)t}\,,\\
e^{-i\left(\mathrm{\Omega}+\omega_{0,\left(\pm\right)}\right)t} & =e^{-i\left({3\Omega}/{2}+\delta\omega_{0,\left(\pm\right)}\right)t}\,, \\
e^{i\left(\mathrm{\Omega}+\omega_{0,\left(\pm\right)}\right)t} & =e^{i\left({3\Omega}/{2}+\delta\omega_{0,\left(\pm\right)}\right)t}\,,\\
e^{-i\left(\mathrm{\Omega}-\omega_{0,\left(\pm\right)}\right)t} & =e^{-i\left({\Omega}/{2}-\delta\omega_{0,\left(\pm\right)}\right)t}\,.
\end{align*}
 By time-averaging upon several cycles, all synchronous terms may be grouped in the vectorial-differential form of Eq.\ \eqref{CWE}, provided that $\delta\omega_{0,\left(\pm\right)}\approx0$.

\section{Solutions of the coupled-wave equations}
\label{Solutions of the coupled-wave equations}
Starting from the system in Eq.\ \eqref{CWE}, it is convenient to rotate the fields into a new basis,
\begin{equation}\label{Kogelnik transformation}
\left[\tilde{y}_{\left(\pm\right)}^t\right]^{\pm}=e^{\mp i\bar{\omega}_{\left(\pm\right)}t}\left[y_{\left(\pm\right)}^t\right]^{\pm}\,,
\end{equation}
which will transform the characteristic matrix in Eq.\ \eqref{Characteristic Matrix} into the Hermitian (for real parameters) and traceless
\begin{equation}\label{Characteristic Matrix Transformed}
    {\bf \tilde{M}}_{\left(\pm\right)}= \left(\begin{matrix}0&i\chi_{\left(\pm\right)}e^{-i2\delta\tilde{\omega}_{\left(\pm\right)} t}\\-i\chi_{\left(\pm\right)}e^{i2\delta\tilde{\omega}_{\left(\pm\right)} t}&0\\\end{matrix}\right)\,.
\end{equation}

\par The coupled-wave equations corresponding to the transformed fields of Eq.\ \eqref{Kogelnik transformation} have analytic solutions in a closed form (cf.\ Eq.\ (3) of \cite{McCall2000}), with an $\rm SU\left(1,1\right)$ underlying symmetry. For $t_0=0$, they can be cast as
\begin{equation}\label{CWE Solutions Transformed}
    {\bf \tilde{A}}_{\left(\pm\right)}\left(t\right)={\bf \tilde{S}}_{\left(\pm\right)}\left(t\right)\cdot{\bf \tilde{A}_{\left(\pm\right)}}\left(0\right)\,,
\end{equation}
where ${\bf \tilde{A}}_{\left(\pm\right)}=\left(\begin{matrix}\left[\tilde{y}_{\left(\pm\right)}^t\right]^+&\left[\tilde{y}_{\left(\pm\right)}^t\right]^-\\\end{matrix}\right)^{\rm T}$ and the components of the transfer matrices are
\begin{equation*}
    {\bf \tilde{S}}_{\left(\pm\right)}=\left(\begin{matrix}e^{- i\delta\tilde{\omega}_{\left(\pm\right)} t}p^{+}_{\left(\pm\right)}&e^{- i\delta\tilde{\omega}_{\left(\pm\right)} t}q^{+}_{\left(\pm\right)}\\e^{ i\delta\tilde{\omega}_{\left(\pm\right)}t}q^{-}_{\left(\pm\right)}&e^{ i\delta\tilde{\omega}_{\left(\pm\right)} t}p^{-}_{\left(\pm\right)}\\\end{matrix}\right)\,,
\end{equation*}
with $p^{\pm}_{\left(\pm\right)}$ and $q^{\pm}_{\left(\pm\right)}$ being those of Eq.\ \eqref{Definition of p and q}.

\par Having solved the coupled-wave equations for the rotated system, it is straightforward to express the general solution to Eq.\ \eqref{CWE} as per Eq.\ \eqref{CWE Solutions}, where
\begin{equation*}
    {\bf S}_{\left(\pm\right)}\left(t\right)={\bf \Phi}_{\left(\pm\right)}^{-1}\left(t\right)\cdot{\bf \tilde{S}}_{\left(\pm\right)}\left(t\right)\cdot{\bf \Phi}_{\left(\pm\right)}\left(0\right)\,,
\end{equation*}
with ${\bf \Phi}_{\left(\pm\right)}={\rm diag}\left(e^{-i\bar{\omega}_{\left(\pm\right)}t},e^{i\bar{\omega}_{\left(\pm\right)}t}\right)$.

\bibliography{sorsamp.bib}

\end{document}